# CW3E's West-WRF 200-member Ensemble

Luca Delle Monache,[a] Daniel F. Steinhoff,[a] Rachel Weihs,[a] Matthew Simpson,[a] Mohammadvaghef Ghazvinian,[a] Vesta Afzali Gorooh,[a] Kevin M. Lupo,[a] Patrick Mulrooney,[a] Caroline Papadopoulos,[a] and F. Martin Ralph [a]

[a] *Center for Western Weather and Water Extremes, Scripps Institution of Oceanography, University of California, San Diego, La Jolla, California*

*Corresponding author*: Vesta Afzali Gorooh, vafzaligorooh@ucsd.edu.

# ABSTRACT

A 200-member ensemble developed at the Center for Western Weather and Water Extremes based on the Weather Research and Forecast atmospheric model tailored for the prediction of atmospheric rivers and associated heavy-to-extreme precipitation events over the Western US (West-WRF) is presented. The ensemble (WW200En) is generated with initial and boundary conditions from the US National Center for Environmental Prediction's Global Ensemble Forecast System (GEFS) and the European Centre for Medium-Range Weather Forecasts' Ensemble Prediction System (EPS), 100 unique combinations of microphysics, planetary boundary layer, and cumulus schemes, as well as perturbations applied to each of the 200 members based on the stochastic kinetic-energy backscatter scheme. Each member is run with 9-km horizontal increments and 60 vertical levels for a 10-month period spanning two winters. The performance of WW200En is compared to GEFS and EPS for probabilistic forecasts of 24-h precipitation, integrated water vapor transport (IVT), and for several thresholds including high percentiles of the observed climatological distribution. The WW200En precipitation forecast skill is better than GEFS at nearly all thresholds and lead times, and comparable or better than the EPS. For larger rainfall thresholds WW200En typically exhibits the best forecast skill. Additionally, WW200En has a better spread-skill relationship than the global systems, and an improved overall reliability and resolution of the probabilistic prediction. The results for IVT are qualitatively similar to those for precipitation forecasts. A sensitivity analysis of the physics parameterizations and the number of ensemble members provides insights into possible future developments of WW200En.

# 1.     Introduction

Atmospheric rivers (ARs) play a pivotal role in global moisture transport from both tropical and extratropical regions, significantly modulating mid-latitude climates (Zhu and Newell 1998; Ralph et al. 2004, 2005; Knippertz and Wernli 2010; Newman et al. 2012). These ARs transport water vapor over a narrow corridor on average 2000 km long and 400 to 800 km wide, predominantly in a vertical layer approximately 3 km deep from the surface (Ralph et al. 2020; Cobb et al. 2021b). In regions such as California, ARs are integral to the water cycle, contributing as much as half of the yearly rainfall (Guan et al. 2010; Gershunov et al. 2017; Dettinger et al. 2011).

Beneficial outcomes, such as the replenishment of water reservoirs and mitigation of wildfire risks are often associated with ARs of weak to moderate intensity (Ralph et al. 2019). Conversely, strong ARs or prolonged and repetitive AR events (Fish et al. 2019) can lead to severe hazards, including localized debris flows (Mcguire et al. 2021; Oakley et al. 2018) and flooding (Corringham et al. 2019). Impactful ARs can also be associated with extreme wind (e.g., Waliser and Guan 2017), resulting in significant damages to properties and infrastructure, particularly in areas with complex topography. Extreme events associated with ARs have been responsible for the majority of insurance claims in the Western US over the past four decades (Ralph et al. 2021). Therefore, accurate prediction of ARs is crucial for managing their potentially hazardous impacts (Dettinger et al. 2011). For example, Forecast Informed Reservoir Operations (FIRO) is an initiative that leverages improved forecasts of AR-related heavy-to-extreme precipitation events to enhance water management and flood hazard mitigation in the western United States (Talbot et al. 2017; Jasperse et al. 2020; Jasperse 2015). The FIRO requires accurate numerical weather predictions (NWP) of ARs to maximize its success, highlighting the need for reliable forecasting systems that can provide early warnings and awareness of significant events (Lavers et al. 2024). The development of observational campaigns like the Atmospheric River Reconnaissance (AR Recon; Ralph et al. 2020) further supports FIRO by improving forecasts of landfalling ARs and their impacts through enhanced observations across the northeast Pacific (e.g., Stone et al. 2020; Zheng et al. 2021; Sun et al. 2022; Dehaan et al. 2023; Lord et al. 2023a,b).

Recent studies have highlighted the variability in forecast skill for both synoptic-scale atmospheric flow and precipitation over California, depending on the prevailing flow regime over the North Pacific. Specifically, forecasts are less accurate in flow regimes featuring a wavy, blocked North Pacific jet stream compared to those with a west-east-oriented jet stream (Moore et al. 2021). This variability in forecast skill relates to the more significant errors observed in forecasts of blocked regimes (Moore 2023). Those are some of the reasons why forecasting extreme weather events, particularly the intensity, structure, and position of ARs over the West U.S. remain challenging. Nardi (2018) and Dehaan et al. (2021) analyzed the performance of global and regional modeling systems in predicting ARs. They found errors ranging from 100 to 350 km in the AR landfall position and 40 to 120 km m$^{-1}$ s$^{-1}$ for integrated water vapor transport (IVT), a measure of AR intensity, depending on the forecast lead time, IVT threshold, and the modeling system evaluated. The skill of the prediction is affected by inaccuracies in the initial conditions, approximation in the numerical procedures and physical packages, and the chaotic nature of the atmosphere that limits its predictability (Lorenz 1963). The latter is best addressed with the generation of ensemble predictions, to capture the range of possible solutions and their evolution in space and time (Epstein 1969a; Leith 1974) while accounting for the above sources of error. A variety of approaches can be found in the literature that are designed to capture the uncertainties stemming from initial conditions (Toth and Kalnay 1997; Buizza and Palmer 1995; Molteni et al. 1996; Johnson and Wang 2016; Raynaud and Bouttier 2016; Schwartz et al. 2021), numerics (Thomas et al. 2002; Berner et al. 2009), and physics (Buizza et al. 1999; Stensrud et al. 2000; Teixiera and Reynolds 2008; Hacker et al. 2011; Bouttier et al. 2012; Jankov et al. 2017; Lupo et al. 2020; Zhou et al. 2022; Kober and Craig 2016; McTaggart-Cowan et al. 2022a,b). Ensembles have improved decision-making in water management and emergency preparedness (Yang et al. 2020; Ndione et al. 2020; Song et al. 2024).

In this paper, we present a regional (i.e., limited area) ensemble specifically designed for the prediction of extreme events associated with ARs. This system is based on the Center for Western Weather and Water Extremes (CW3E) Weather Research and Forecast atmospheric model (Skamarock et al. 2019) tailored for the prediction of ARs over the Western US (West-WRF; Martin et al. 2018). The West-WRF ensemble (hereafter referred to as WW200En) includes 200 members, each run at 9-km horizontal grid spacing. The hypothesis is that, everything else equal, a properly designed ensemble with a larger number of members better samples the right tail of the predictive distribution, which corresponds to heavy-to-extreme precipitation events and high-intensity ARs. In this study, these events are identified by evaluating the forecasts for the prediction of precipitation above several thresholds, up to 90 mm, and percentiles, up to the 99$^{th}$. The WW200En performance is compared to state-of-the-science systems, such as the US National Center for Environmental Prediction (NCEP)'s Global Ensemble Forecast System (GEFS) and the European Centre for Medium-Range Weather Forecasts (ECMWF)'s Ensemble Prediction System (EPS). Important attributes of probabilistic forecasts such as reliability, resolution, and spread-skill relationship are evaluated for the three systems over the Western US and 10 months of data (over which all systems are available), to assess the overall performance and the ability to predict extreme events. We also present a sensitivity analysis of important aspects of the WW200En ensemble design, as well as its performance for two case studies associated with the landfall of impactful ARs.

Section 2 describes the study area and data sources. Details regarding the WW200En design are provided in section 3. Sections 4 and 5 describe the verification metrics, provide the results, and present a sensitivity analysis of WW200En. Discussion and concluding remarks are highlighted in section 6.

## 2. Data

This section describes the data sets utilized for the development and testing of WW200En. Those include the global and regional NWP systems and the ground-truth data sets for precipitation and IVT. Data was collected over the study period from 1 December 2021 to 31 March 2022 and from 1 October 2022 to 31 March 2023 and over the WW200En computational domain (Fig. 1).

### a. *Global Ensemble Forecast Models*

In addition to the WW200En, precipitation forecast skill is evaluated for ensemble forecasts from NCEP's Global Forecast System (GFS) (Environmental Modeling Center (EMC) 2024) and ECMWF's EPS (ECMWF 2021) global ensemble prediction systems. Moreover, initial and boundary conditions for the WW200En are derived from these global operational systems (described in section 3.2). Deployed in October 2020, GEFS version 12 consists of 31 members based on GFS version 15.1 (EMC 2019) utilizing the Finite-Volume Cubed-Sphere dynamical core (FV3; Geophysical Fluid Dynamics Laboratory) and is run at ~25 km horizontal grid spacing with 64 vertical levels (EMC 2024). Initial condition uncertainty in the GEFS is represented using perturbations from Global Data Assimilation System (GDAS) 80-member Ensemble Kalman Filter (EnKF), and model uncertainty is represented by both the stochastic perturbed parameterization tendencies (SPPT) and stochastic kinetic energy backscatter (SKEB) schemes (EMC 2024). ECMWF's IFS CY47R3 was operational from September 2021 to June 2023 (encompassing the 10-month period spanning two cool seasons examined herein) and included a 51-member ensemble forecast system with ~18 km horizontal grid spacing and 91 vertical levels, with initial condition uncertainty represented by perturbations based on an ensemble of data assimilations (EDA) and initial-time leading singular vectors, and model uncertainty represented using SPPT (ECMWF 2021). Verification metrics from GEFS and EPS forecasts were computed using operational model output with horizontal grid spacings of 0.25° and 18 km, respectively.

b. *Ground-truth Data Sets*

1) PRECIPITATION

In this study, we use 24-h total precipitation rates derived from the Parameter Elevation Regression on Independent Slopes Model (PRISM; Daly et al. 2008) as a reference dataset, featuring a spatial grid spacing of approximately 4 km and valid every day at 12:00 UTC. The forecasts were budget interpolated to align with the grid points of the PRISM analyses. For precipitation, the verification period is 10 months across the two water years, from December to March 2021-2022 and October to March 2022-2023 over the area indicated in Fig. 1.

2) IVT

The ground-truth estimates of IVT are derived from dropsondes collected during AR Recon (Ralph et al. 2020). These measurements allow the evaluation of the performance of the prediction systems in the presence of ARs, which is the focus of this study, and also to assess the value of a higher resolution ensemble as WW200En (9 km) when compared to coarser systems such as GEFS (25 km) and EPS (18 km). Such evaluation would have not been possible if a reanalysis data set would have been used to generate ground-truth estimates of IVT, as for example, the ECMWF's Reanalysis v5 (ERA5) as suggested by (Cobb et al. 2021a), although in that case more samples would have been available to compute the performance metrics. Additionally, since ERA5 is generated with a dynamical model and data assimilation system very similar to the EPS, that could have skewed the comparison presented in this study towards the ECMWF's ensemble.

The dropsondes were deployed from aircraft and provide vertical profiles of temperature, humidity, wind and pressure as they descend through the atmosphere. Their deployment in AR Recon missions has been shown to benefit the initialization of NWPs significantly and, consequently the prediction of high-impact precipitation events associated with ARs (Stone et al. 2020; Zheng et al. 2021, 2024; Sun et al. 2022; Dehaan et al. 2023; Lord et al. 2023a,b). The IVT estimates are calculated from 1218 vertical profiles from dropsondes, available from 4 November 2022 to 14 March 2023 and over the area illustrated in Fig. 1.

## 3. 200-member West-WRF Ensemble Design

The underlying dynamical model of WW200En is CW3E's West-WRF, a version of the WRF model (Skamarock et al. 2019) that is tailored for the prediction of extreme events associated with ARs over Northeastern Pacific Ocean and the Western US (Martin et al. 2018). The West-WRF system has been shown to skillfully predict atmospheric river activity and precipitation patterns to support better water management decisions in the West (Martin et al. 2018; Dehaan et al. 2021; Cobb et al. 2023). The West-WRF has been used to generate near-real-time deterministic forecasts during the boreal winter season of each year since 2015 and for the production of a 34-year reforecast dataset (Cobb et al. 2023) for studies on climatological patterns of precipitation, atmospheric river activity, and the development of new post-processing and machine learning techniques (Chapman et al. 2022; Badrinath et al. 2023; Hu et al. 2023). Previously, the CW3E West-WRF system consisted of two deterministic forecast simulations that ran during the cool season (December-March) using initial conditions from GFS and ECMWF.

The WW200En is based on WRF version 4.5.1 with horizontal grid spacing of 9 km, and 828 by 570 east-west and north-south grid cells, respectively. Sixty terrain following eta levels and a model top of 10 hPa (approximately 28-32 km above sea level) comprise the ensemble vertical grid configuration. Ensemble eta spacing as a function of height is generated using the WRF default vertical grid methodology. Forecasts are initialized at 00Z and run out through 7 days.

The current WW200En domain geographic coverage (Fig. 1) is the result of years of development and reflects the need to capture areas to the west side of the domain where ARs form and then evolve, which is beneficial to the skill in predicting these atmospheric phenomena and the associated precipitation at landfall as well as critical for the forecast information necessary for the execution of the CW3E-led AR Recon. Also, the ensemble domain was expanded to the east to capture the evolution of storm systems along the Rocky Mountains that can impact weather over the Western US.

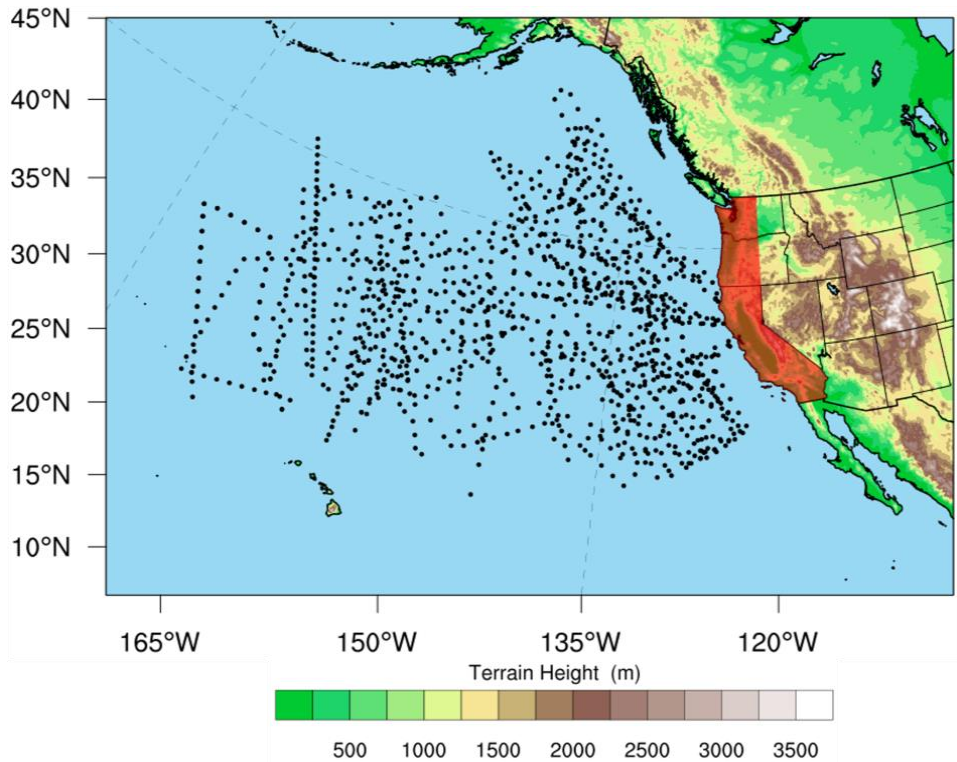

Fig 1. Geographic coverage of the 9 km horizontal grid spacing WW200En numerical domain for Water Years 2022 and 2023. Dropsonde locations during the period of study are shown with black dots. Red translucent shading shows the PRISM verification region.

The WW200En is designed to accurately forecast ARs and the associated extreme precipitation over the western U.S., while reliably quantifying the uncertainty associated with these predictions. To that end, the design is based on the implementation of different perturbation strategies to capture the uncertainty stemming from its main sources in an NWP system: the initial conditions, the numerical procedures, and the physics packages (Fig. 2). Next, the rationale of the chosen number of ensemble members is explained, followed by a description of the perturbations utilized to generate WW200En.

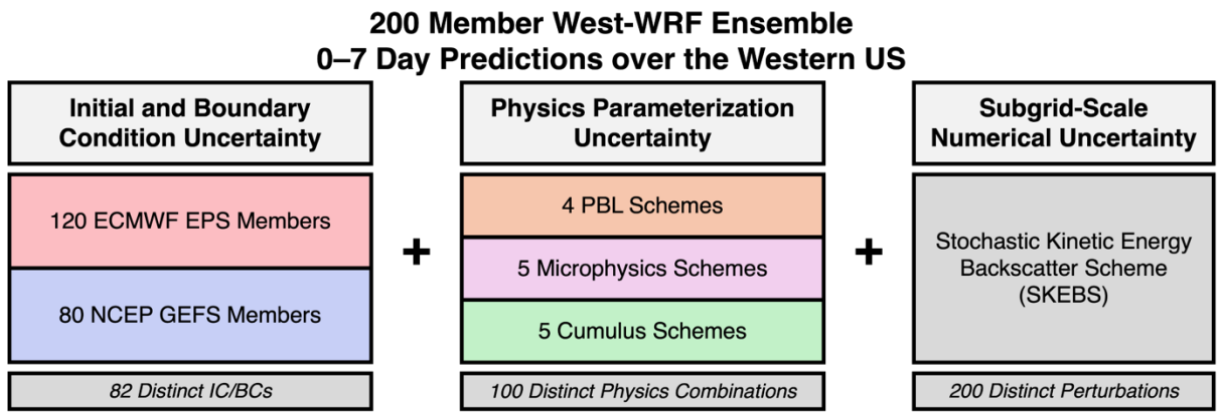

Fig 2. Schematic showing the generation of 200 unique ensemble members by utilizing GEFS and ECMWF ensemble initial conditions, multi-physics combinations, and SKEB perturbations.

*a. Number of ensemble members*

The number of members in an ensemble is typically codependent on the ability to capture the uncertainty of the prediction of interest and the limit posed by the available computational resources. To that end, CW3E had exclusive access to the *Comet* supercomputer (San Diego Supercomputer Center 2021a, 2021b) during the period of study[1]. An ensemble size of 200 members was chosen for several reasons; it far exceeds the current number of available ensemble members from prominent forecast centers (e.g., NCEP's GEFS 31-member and ECMWF's EPS 51-member ensembles) with the expectation that it can better resolve the tails of the atmospheric state distribution. The tails represent the weather extremes − often associated with the highly variable frequency and intensity of atmospheric rivers, which the West-WRF model is designed to target. It was also chosen as a target number to adequately span the prediction uncertainty with the methods described in the next sections, using a large portion of Comet while leaving the rest of the supercomputer to other ongoing research activities at CW3E.

*b. Initial and lateral boundary condition uncertainty*

Initial and boundary condition uncertainty is sampled in the WW200En by utilizing perturbed analysis states from the GEFS and EPS operational ensembles[2]. Perturbed initial conditions of the GEFS system are generated from a 6-h ensemble Kalman filter forecast (Zhou et al. 2022). Thirteen GEFS members are used twice, and 18 members used three times, leading to a total of 80 WW200En members. The suite of analysis states of the EPS system is generated by a combination of perturbations from an ensemble of data assimilations and constructed from the leading singular vectors (ECMWF 2023). Thirty-three EPS perturbed initial and boundary conditions are used at least twice, and 18 members are used three times, leading to a total of 120 WW200En members based on the EPS. The additional members are used to sample the full range of physics parameterizations to be discussed in section 3.4; these members are randomly sampled from the EPS and GEFS. The larger sampling of the EPS members reflects the fact that there are more members for the EPS than GEFS.

---

[1] CW3E retained exclusive use of Comet following its retirement as an XSEDE resource in 2021 through March 2024. Of the available 1994 nodes, 1200 were used to generate WW200En near-real time forecasts. Each node had 24 cores with a 2.5 GHz clock, 128 GB of DRAM, and 320 GB of SSD memory. Members of the WW200En each took ~3–4 hours of wallclock time, which varied based on the computational cost of each parameterization scheme, with 16 nodes for each member.

[2] The native horizontal grid spacing of the WW200En input data was 18 km for EPS and 0.5° for GEFS. While 0.25° GEFS data is available and used for computing verification metrics, the limited selection of parameters at finer horizontal resolution (NCEP 2024) are insufficient for running WRF.

*c. Numerical procedure uncertainty*

The incorrect representation of the upscale energy transfer from smaller to larger scales is another source of model error that should be accounted for when constructing an ensemble. The correct representation of these processes may be limited by numerical procedures designed to generate a stable integration at the cost of a smoother solution. The stochastic kinetic-energy backscatter scheme (SKEB; Berner et al. 2009, 2008; Marzban et al. 2019) was proposed to stochastically perturb a forecast using components of the rotational wind components, u and v, and potential temperature. The amplitude of the wind perturbations is proportional to the square root of the kinetic-energy backscatter rate and the amplitude of the temperature perturbations are proportional to the potential energy backscatter rate. The SKEB scheme is applied to every member with a unique seed used to generate the perturbation for each member. No tuning was applied to the default SKEB scale parameters.

*d. Physics parameterization uncertainty*

The WRF model can simulate several sub-grid physical processes including cumulus convection development, planetary boundary layer (PBL) processes and surface heat transfer, and microscale dynamics within the atmosphere. Several packages are available for the model to choose from for each process based on the targeted weather phenomena and application. To account for uncertainties in the methods used to parameterize these key physical processes, members of the WW200En are assigned varying combinations of physics packages. Table 1 shows the chosen physics packages for each of the sub-grid processes: microphysics (4 schemes), cumulus (5 schemes), and PBL (4 schemes). Combined, these yield a total of 100 distinct combinations, each repeated twice across the 200 members of WW200En. In general, the schemes included in the WW200En configuration were chosen to sample across multiple conceptual approaches within each parameterization category. For example, for the PBL, we included schemes based on local (MYJ and MYNN), non-local (YSU), and a combination of local and non-local (ACM2) closures. These schemes were additionally bounded by testing of compatibility and computational efficiency required for near-real time forecasting. Results on performance of each parameterization category and combinations of parameterizations is given in section 5d.1.

Table 1. List of physics parameterizations utilized across the 200-member West-WRF ensemble and options used for each parameterization. References for each parameterization scheme are given in parenthesis.

| Microphysics | Cumulus | Planetary Boundary Layer |
|---|---|---|
| Morrison (Morrison et al. 2009) | Grell-Freitas (Arakawa 2004; Grell and Freitas 2014) | Mellor-Yamada-Janjic (MYJ; Janjić 1994; Mesinger 1993) |
| Stony Brook University (SBU; Ylin) (Lin and Colle 2011) | Korea Inst. of Atmo. Prediction Systems (KIAPS) simplified Arakawa-Schubert (SAS; Han and Pan 2011; Kwon and Hong 2017) | Mellor-Yamada-Nakanishi-Niino (MYNN; Nakanishi and Niino 2006, 2009; Olson et al. 2019) |
| WRF Single-Moment 6-Class (WSM6; Hong and Lim 2006) | New Tiedtke (Zhang and Wang 2017) | Asymmetric Convective Model, Version 2 (ACM2; Pleim 2007) |
| Purdue Lin (Chen and Sun 2002) | Multi-scale Kain-Fritsch (Zheng et al. 2016; Glotfelty et al. 2019) | Yonsei University (YSU; Hong et al. 2006) |
| Thompson (Thompson et al. 2008) | Betts-Miller-Janjic (BMJ; Janjić 1994) | |

## 4. Verification Metrics

Forecasts of precipitation and IVT are evaluated to analyze the ability of the global and regional ensembles included in this study to predict AR conditions and the associated impacts. For threshold-based evaluations, the Brier Score (Brier 1950),

$$BS = \frac{1}{N}\sum_{i=1}^{N}(F_i(t) - I\{y_i \leq t\})^2$$

is the mean squared error for probability prediction, where *N* is the total number of pairs of forecasts and observations, *t* is threshold value, *F* denotes forecast cumulative distribution function (CDF) and *I* is the step function that takes value 1 if the threshold t exceeds the ith verifying observation and 0 otherwise. In order to assess the performance of specific forecasts in predicting binary events relative to a reference forecast, we use the Brier Skill Score (BSS),

$$BSS = 1 - BS/BS_{ref}$$

The BSS when multiplied by 100%, can be interpreted as the percent improvement (or worsening) of the Brier Score relative to the reference forecast. In section 5, the PRISM monthly climatology is used as the reference forecast for comparisons between the ensemble precipitation forecasts from WW200En, GEFS, and ECMWF EPS, and is derived separately for each grid point and month of the year using the 24-h accumulated the PRISM historical data from that month, and two surrounding months. For IVT, it is not possible to compute climatological percentiles because the IVT dropsonde record is too short for that purpose. For comparisons between different subsets of the WW200En, the full WW200En is used as the reference forecast. We note that in this study, we consider precipitation and IVT thresholds that are typically observed across the western U.S. during AR events, i.e., 0.254, 25, 40, 50, 60, 70, 75, 80, and 90 mm for 24-h accumulated precipitation, and 250, 500, and 750 kg m$^{-1}$ s$^{-1}$ for IVT.

The Brier score components are also very useful to evaluate two important attributes of probabilistic prediction, *resolution and reliability.* The BS has a third component, *uncertainty*, which depends solely on the sample climatology and is independent of the forecast source (Murphy 1973; Wilks 2011).

$$BS = \text{Reliability} - \text{Resolution} + \text{Uncertainty}$$

$$BS = \frac{1}{N}\sum_{i=1}^{K} N_i \left[F_i(t) - \underline{o}_i(t)\right]^2 - \frac{1}{N}\sum_{i=1}^{K} N_i \left[\underline{o}_i(t) - \underline{o}(t)\right]^2 + \underline{o}(t)\left[1 - \underline{o}(t)\right]$$

where K represents the number of categories to which forecasts are aggregated, $\underline{o}_i(t)$ is the average climatological probability exceeding the threshold t in that category, and $\underline{o}(t)$ is the overall average climatological probability. Resolution represents the ability of a forecast to discriminate between observable future events. Reliability (calibration) refers to statistical compatibility between the ensemble forecast distribution and the observations. An ensemble of forecasts with perfect reliability always predicts probabilities that match the observed relative frequencies of a given event (Wilks 2011; Jollife and Stephenson 2012). Probabilistic forecasts with lower reliability and higher resolution values are desirable.

Reliability diagrams can be used to understand the ensemble calibration graphically and for a range of thresholds (e.g., Brocker and Smith 2007; Wilks 2011). For a specified threshold, the reliability diagram compares the average forecast probability to the observed relative frequency for various forecast probability bins. In this study, forecasts are categorized using twelve, equally-spaced probability bins between 0.0 and 1.0. In a perfectly reliable ensemble, the reliability curve should lie along the 1:1 line (e.g., for a forecast probability of 30%, the event should have a relative frequency of 30%). Forecast frequency of usage (sharpness) curves are included to investigate the frequency of forecast issuance probabilities for different categories. Higher forecast frequencies for probabilities close to either 0 or 100% are preferred.

The Continuous Ranked Probability Score (CRPS; Matheson and Winkler 1976; Gneiting and Raftery 2007;) is a common metric to estimate the overall probabilistic predictive performance. The CRPS integrates squared differences between the CDF of probabilistic forecast and the verifying observation and can be interpreted as the integral of Brier score.

$$\text{CRPS} = \frac{1}{N}\sum_{i=1}^{N} \int_{-\infty}^{\infty} [F_i(x) - I\{y_i \leq x\}]^2 dx$$

where F is the CDF of the ensemble forecast, and y is the verifying observation. The step function I takes the value of 1 if $y \leq x$ and 0 otherwise. Similar to the BSS, the continuous ranked probability skill score (CRPSS) is computed as follows:

$$CRPSS = 1 - CRPS/CRPS_{ref}$$

where $CRPS_{ref}$ is the CRPS of a reference forecast, similar to the $BS_{ref}$ in the calculation of the BSS.

Weighted scoring rules can be computed when the focus is on the right tail of forecast distribution (more impactful values of interest). Following Lerch et al. (2017) and Gneiting and Ranjan (2011) the threshold CRPS (twCRPS) is computed as

$$\text{twCRPS} = \frac{1}{N}\sum_{i=1}^{N} \int_{-\infty}^{\infty} \omega(x)[F_i(x) - I\{y_i \leq x\}] dx$$

where $\omega$ is a non-negative weight function. When $\omega(x) = 1$, the twCRPS simplifies to the CRPS. Typically, $\omega$ is a step function of the form $\omega(x) = I\{x \geq t\}$ for some high threshold t, with a value of 1 for values larger than t and 0 elsewhere. The threshold can be chosen as hard thresholds (fixed values) or determined by the distribution of observed data such as using larger percentiles (e.g., > 90th percentile) specific for each grid point and month of the year. We chose to use the latter values in our computations to account for the fact that the climatological distributions can vary dramatically across different locations over the spatial domain considered (see Fig. 1; e.g., consider a grid point over the Sierras vs a grid point over the semi-arid region in southeast California). Skill scores for the twCRPS are computed similarly to CRPSS and BSS as indicated earlier.

Another important attribute of an ensemble-based probabilistic prediction, is the ensemble statistical consistency and its ability to capture flow-dependent uncertainty. This can be estimated by compiling binned spread-skill diagrams (Van Den Dool 1989; Delle Monache et al. 2013; Chapman et al. 2022; Ghazvinian et al. 2024) where the ensemble spread (i.e., standard deviation of the ensemble) is compared to the root mean square error (RMSE) of the ensemble mean. A perfect spread-skill relationship aligns with the 1:1 line.

We finally evaluate relative operating characteristics (ROC) diagrams and associated ROC skill scores (ROCSS) to better understand the forecasts' ability to discriminate between events and non-events (Jollife and Stephenson 2012; Junk et al. 2015). The ROC diagrams are generated by plotting false alarm rate (probability of false detection; abscissa), against hit rate (probability of detection; ordinate) for exceedance thresholds levels between 0% and 100%. We further compute the area under the ROC curve (AUC) and compute the ROC skill score (ROCSS) as follows:

$$\text{ROCSS} = \frac{A - 0.5}{1 - 0.5} = 2A - 1$$

where the AUC is denoted by A. For a perfect forecast, A = 1 and ROCSS = 1. For the sample climatology, A = 0.5 and ROCSS = 0 (ROCSS < 0 for a forecast worse than climatology). We compute ROC diagrams for the same binary events as shown for BSS and reliability diagrams for both IVT and precipitation forecasts.

The statistical significance of verification score differences (e.g., CRPS, BS) between ensemble systems is evaluated through bootstrap resampling using 1000 samples with replacement (e.g., Brocker and Smith 2007; Junk et al. 2015). Statistical significance is determined at the 90% bootstrap confidence interval (i.e., using the 5th and 95th percentile) and is evaluated through visual assessment of the confidence intervals overlayed on curves in time-series plots, as well as with colored dots at the top of the figure, indicating statistically significant differences between a particular pair of forecasts.

# 5. Results

The performance of the ensemble forecast systems described herein (i.e., GEFS, EPS, and WW200En) is evaluated for the precipitation and IVT prediction of both common and rare events. Individual case studies are also examined to provide a detailed analysis of probabilistic predictions for specific events, and a sensitivity analysis of key aspects of the proposed ensemble design is presented at the end of this section.

### a. *Precipitation*

As indicated in Figure 1, verification metrics for 24-hour accumulated precipitation forecasts over the western U.S. are computed over a region extending from the Pacific coastline to several hundred miles inland, encompassing all of California and the western portions of Oregon and Washington State and representing the area over which the West-WRF has been most refined and optimized since its inception. The WW200En compares favorably to the global systems from which it is built, with a CRPSS score higher than GEFS at every lead time and similar or slightly higher in day 1 and 2 than EPS and lower than that in day 4 and day 5. The differences between EPS and WW200En are statistically significant at the aforementioned lead times but not at day 3 and day 6 (Fig. 3).

To assess the performance of the ensembles for different precipitation intensities, the BSS is shown in Figure 4 for nine 24-h precipitation thresholds, starting at 0.254 mm (a proxy for yes/no precipitation) up to 90 mm (associated with heavy precipitation events). For the lowest threshold, WW200En performs well with respect to the global systems, exhibiting a higher BSS at each lead time during forecast days 1–6 (Fig. 4a). The improvement in low-threshold BSS with respect to the global systems generally diminishes with increasing forecast lead time, particularly in comparison to EPS (Fig. 4a). Similar results are observed for higher precipitation thresholds, such that WW200En ensemble has a higher or comparable BSS to EPS, and often a higher BSS than the GEFS, except for 60- and 70-mm thresholds at 2-day lead times (Fig. 4). Overall, these differences yield ~0.5–2-days of additional precipitation predictability, depending on the chosen threshold, for the regional CW3E regional ensemble compared to the two global systems, and often are statistically significant between WW200En and GEFS (except at day 2), and to a lesser extent between WW200En and EPS.

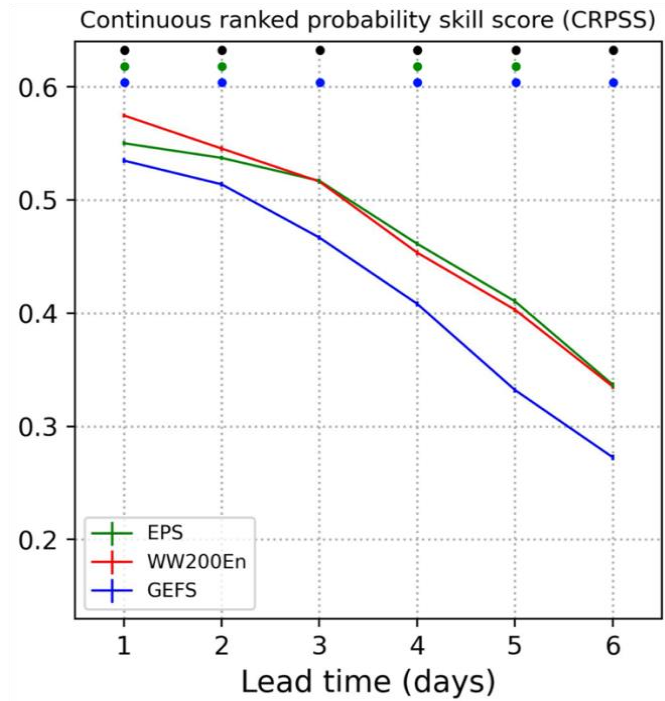

Fig 3. Continuous ranked probability skill score (CRPSS) for 24-hour precipitation over the PRISM verification region (red shading in Fig. 1). Bootstrap 90% confidence intervals are plotted for each ensemble system as vertical bars at each forecast day. Results are aggregated over all available grid points covering the verification area and all verification days as a function of lead time. The score is calculated with respect to climatology, and PRISM gridded observations are the ground-truth. As a visual aid, markers are placed along the upper x-axis denoting lead times where the differences between ensembles are statistically significant at the 90% bootstrap confidence interval. Black markers refer to the difference between EPS and GEFS, blue markers refer to the difference between WW200En and GEFS, and green markers refer to the difference between WW200En and EPS.

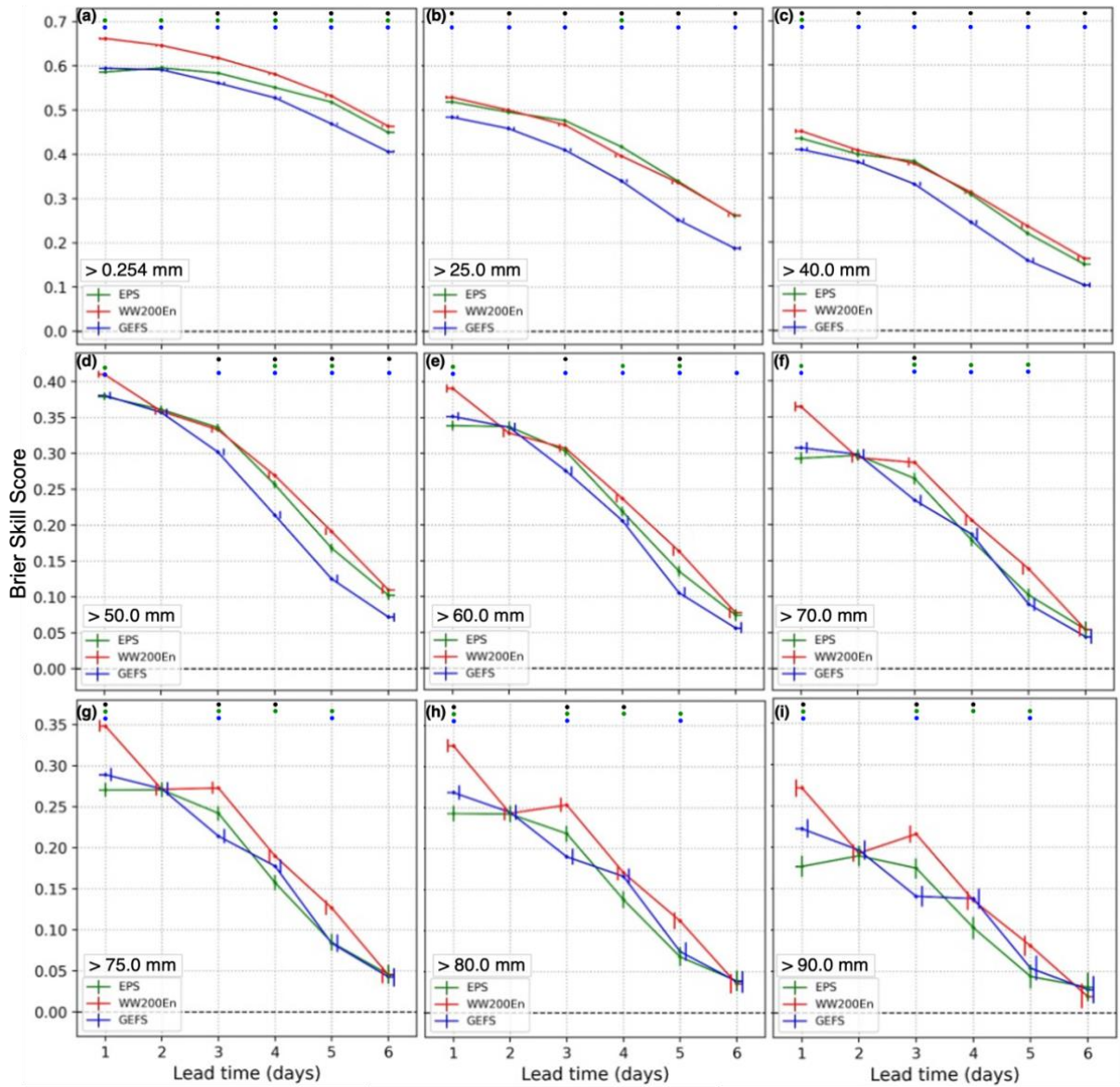

Fig 4. Brier skill scores (BSS) from all forecasts of 24-h accumulated precipitation for the following thresholds: (a) > 0.254 mm, (b) > 25 mm, (c) > 40 mm, (d) > 50 mm, (e) > 60 mm, (f) > 70mm, (g) > 75 mm, (h) > 80 mm, and (i) > 90 mm. Bootstrap 90% confidence intervals are plotted for each ensemble system as vertical bars at each forecast day. Results are aggregated over all the grid points covering the verification area and all verification days as a function of lead time. The score is calculated with respect to climatology, and PRISM gridded observations are the ground-truth. As a visual aid, a small offset was applied to the x-coordinates of the GEFS and WW200En bootstrap confidence intervals, and markers are placed along the upper x-axis denoting lead times where the differences between ensembles are statistically significant at the 90% confidence interval. Black markers refer to the difference between EPS and GEFS, blue markers refer to the difference between WW200En and GEFS, and green markers refer to the difference between WW200En and EPS.

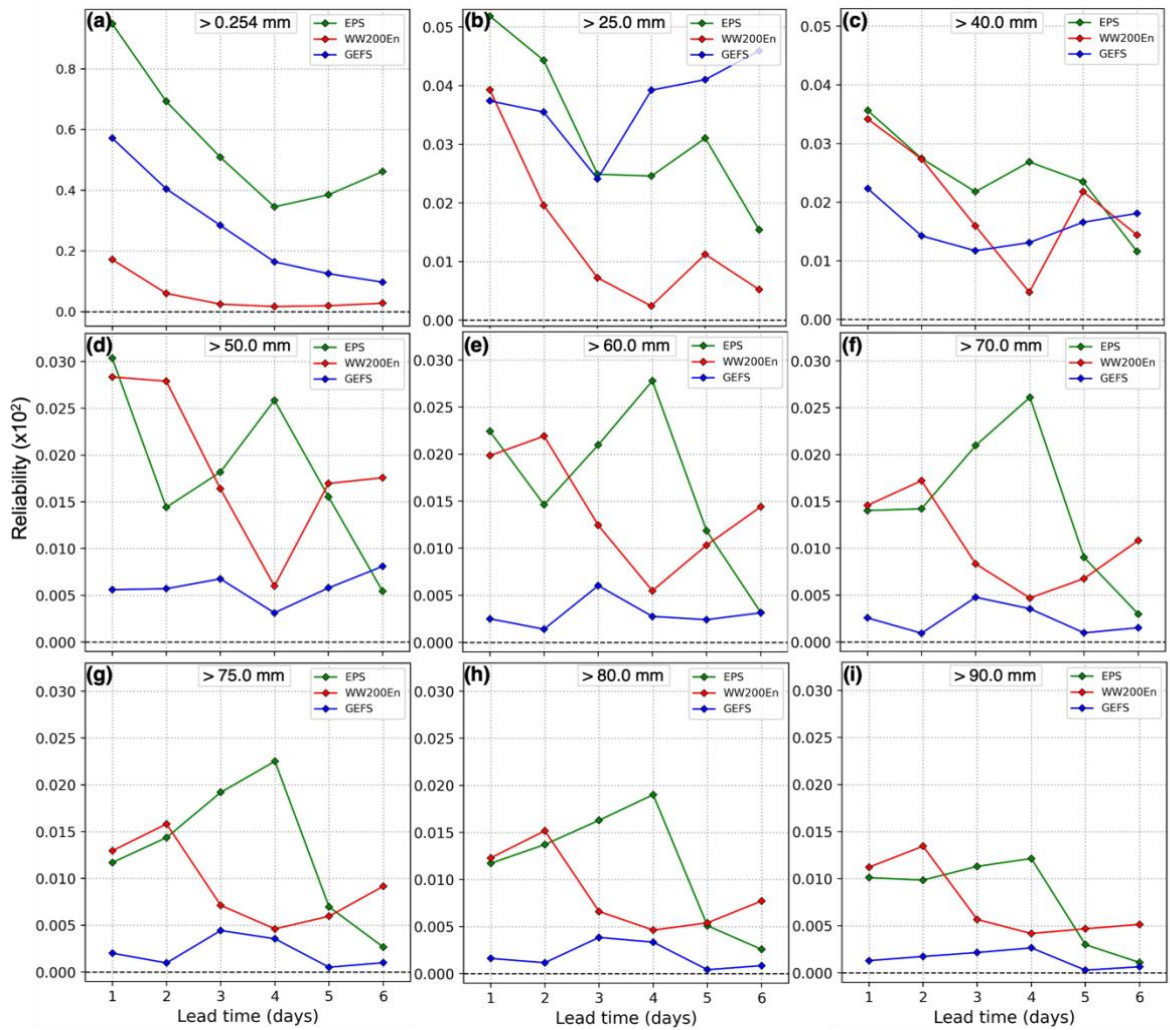

Fig 5. As in Figure 4, but for the Brier score reliability component (lower is better). Values are scaled by 100 for clarity.

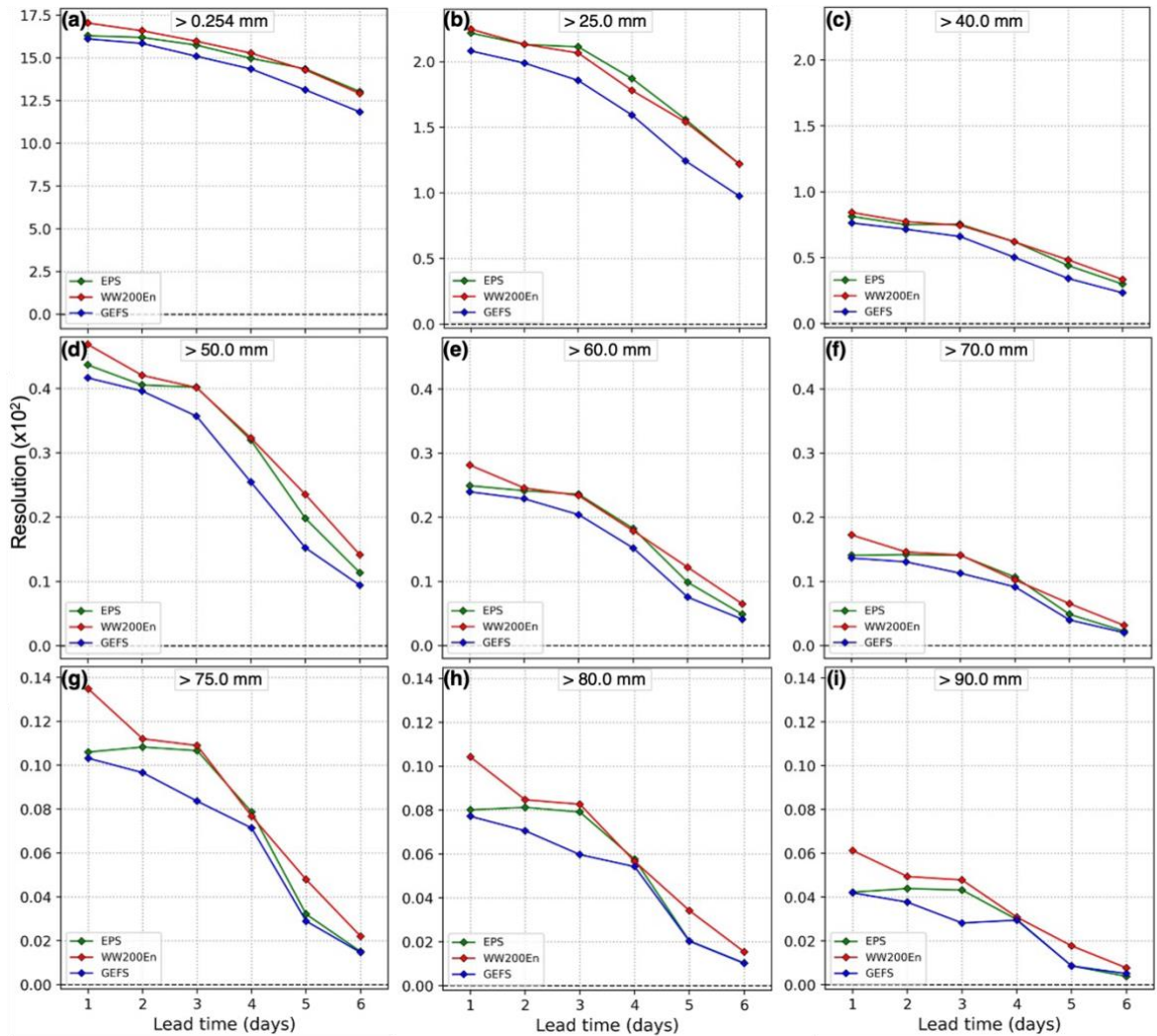

Fig 6. As in Figure 4, but for the Brier score resolution component (larger is better). Values are scaled by 100 for clarity.

The reliability scores for each ensemble system considered in this study are near zero, indicative of well-calibrated precipitation forecasts (Fig. 5). The WW200En has the lowest (best) reliability scores over all forecast lead times for the lowest two thresholds (0.254 and 25 mm; Figs. 5a and 5b, respectively). The GEFS is characterized by the best reliability scores at each forecast lead time for each 24-h precipitation threshold ≥ 50 mm (Fig. 5). For these higher thresholds, EPS has somewhat poorer reliability, particularly at forecast days 3 and 4 (Fig. 5). Causes of this behavior in EPS ensemble are not presently identified.

For resolution, WW200En and EPS have higher (better) scores than GEFS at nearly all forecast lead times and thresholds (Fig. 6). The WW200En generally has the largest resolution score for the highest 24-h precipitation thresholds (75, 80, and 90 mm; Figs. 6g–i) at all lead times excluding forecast day 4.

File generated with AMS Word template 2.0

In the binned spread–skill framework, each of the three ensemble systems examined herein are associated with spread-deficient (i.e., underdispersive) precipitation forecasts at 1 and 3 day forecast lead times (Figs. 7a and b), with WW200En exhibiting spread more representative of the mean error by day 6 (Fig. 7c). In this framework, this is visually assessed by the position of the spread–skill curves to the left of the 1:1 line in Figure 7, such that for a given forecast spread bin, the ensemble mean error is *larger* than the spread. The WW200En is the closest to the 1:1 line, or "perfect forecast," for 1, 3, and 6-day lead times (Fig. 7), whereas GEFS and EPS perform similarly to each other, but are more underdispersive. For 6-day forecasts, the spread–skill curve of the 200-member ensemble is nearly perfect, overlapping or very close to the 1:1 line (Fig. 7c). These results indicate that the spread of WW200En (as well as the global systems) does not grow quickly enough, taking on average 6 days for WW200En's forecast uncertainty to closely represent the error of the ensemble mean.

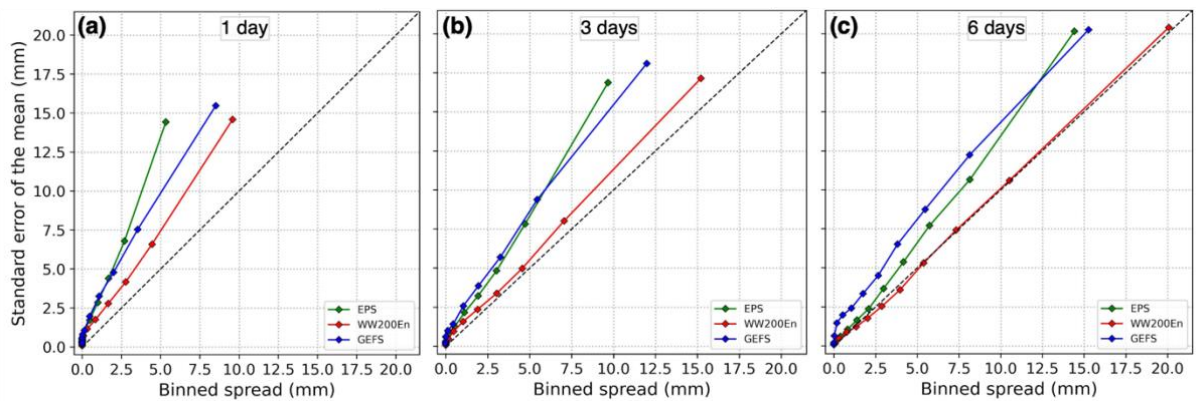

Fig 7. Binned spread–skill diagrams for 24h accumulated precipitation forecasts and for lead times of 1 day (left; +12 to +36h), 3 days (middle; +60 to +84 h), and 6 days (right; +132 to +156h).

Reliability diagrams and ROC curves are examined at forecast day 3 for 0.254, 60, and 90 mm 24-hour precipitation thresholds to further evaluate the relationship between probabilistic forecasts and observed events. For each ensemble system, both false alarm rates and hit rates generally decrease with increasing precipitation threshold (Fig. 8a–c), with false alarm rates only exceeding 0.2 for the lowest precipitation threshold (Fig. 8a). Total ROCSS values also decrease for each ensemble system with increasing precipitation threshold, consistent with smaller hit rates for larger precipitation totals (Fig. 8a–c). At all three 24-hour precipitation thresholds, the ROCSS of the WW200En is larger than for the two global systems, with the skill score difference increasing with increasing precipitation thresholds (Fig. 8). This is consistent with the behavior of reliability diagrams for the same three precipitation thresholds (Figs. 8d–f), such that for higher thresholds, the WW200En 3-day reliability curve remains to the right of the 1:1 line (i.e., a slight over-forecasting bias in all probability bins), and the reliability curves for the global systems, most notably the EPS, are left of the 1:1 line (i.e., an under-forecasting bias in all probability bins). Reliability diagrams are also consistent with their Brier Score components (i.e., reliability and resolution; Figs. 5 and 6, respectively), such that for the lowest precipitation threshold, the WW200En has nearly perfect reliability (c.f., Figs. 5a and 8a), and for the higher precipitation thresholds, the WW200En has improved resolution, as indicated by the relatively higher frequency of large forecast probabilities in the inset sharpness diagrams in Figures 8e and f, particularly compared to GEFS (c.f., Figs. 6e, i and Figs. 8e, f). Overall, while 3-day forecasts from each ensemble system experience a loss of skill for higher precipitation thresholds, forecasts from the WW200En, which are built using the two global systems as initial and boundary conditions, appear to better depict larger precipitation totals compared to GEFS and EPS forecasts.

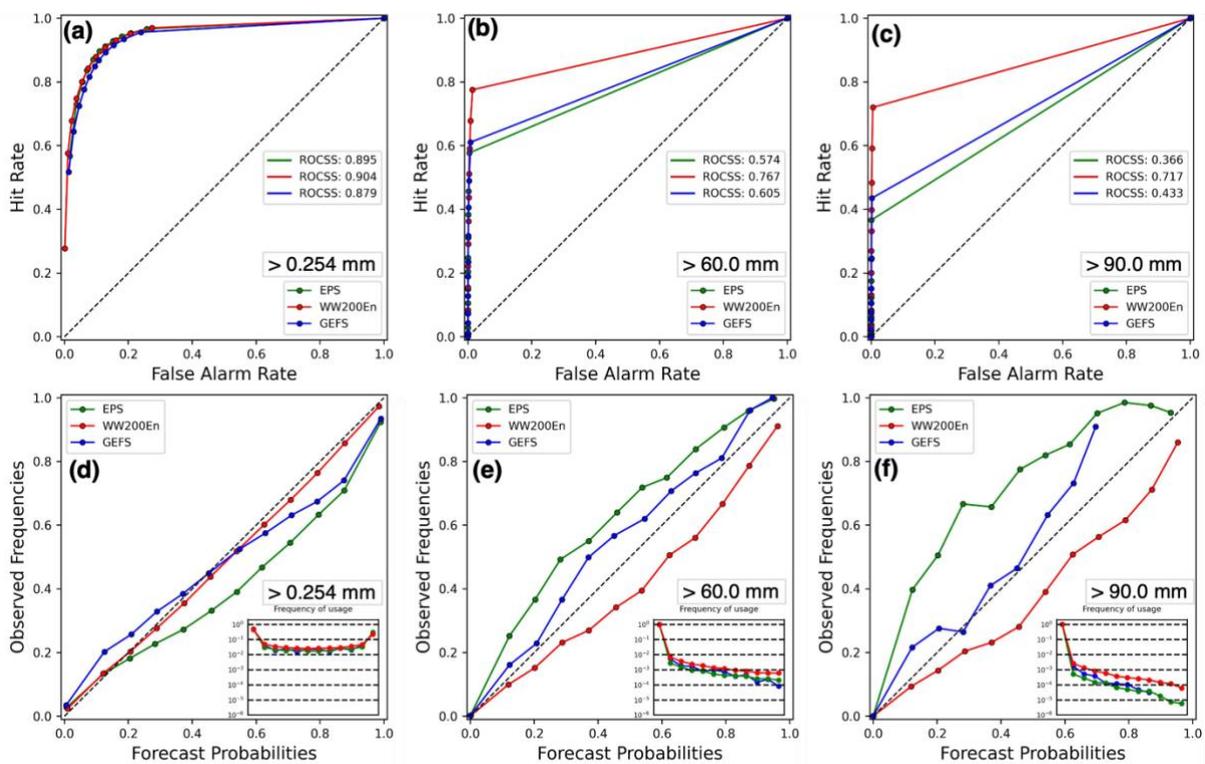

Fig. 8. (a–c) ROC diagrams and ROC skill score (ROCSS; mid-right inset) and (d–f) reliability diagrams and frequency of forecast usage (sharpness diagrams; log10 scale) from all 3-day forecasts of 24-h accumulated precipitation for (a and c) > 0.254 mm, (b and e) > 60 mm, (c and f) > 90 mm precipitation thresholds.

*b. IVT*

To evaluate the performance of the GEFS, EPS, and WW200En in representing the larger-scale processes leading to precipitation over the western U.S., ensemble forecasts of IVT over the North Pacific are verified against 1218 dropsonde observations during the 10-month period of study. Similar to precipitation forecasts, the EPS and WW200En are associated with lower (better) CRPS values than the GEFS for IVT throughout the 6-day forecast period (Fig. 9). While these two ensembles have comparable CRPSs for most of the forecast period, the 200-member ensemble has the lowest CRPS for all forecast lead times except for day 5, with the largest difference during forecast days 2–4 (Fig. 9). The differences between WW200En and EPS with respect to GEFS are significant at days 1, 2, 5, and 6.

Brier scores, representative of the mean-squared probability error, are also the smallest for the 200-member ensemble for three IVT thresholds (Fig. 10; IVT > 250, 500, and 750 kg m$^{-1}$ s$^{-1}$). These IVT values correspond to minimum thresholds used to classify 24-h AR events as "Weak," "Moderate," or "Strong" ARs at a grid point based on the AR scale introduced by Ralph et al. (2019; their Fig. 4). Similar to the IVT CRPS scores, WW200En and EPS have the lowest Brier scores for nearly all forecast days over the three thresholds, performing more similarly to each other than the GEFS (Fig. 10). Moreover, WW200En, having the smallest Brier scores for all forecast lead times except for day 5, has its largest difference with respect to EPS during forecast days 2–4, and the largest difference with respect to GEFS during days 5–6 for each threshold (Fig. 10), although none of these differences are statistically significant. Similarly, comparison of ROC curves and ROCSS for the three IVT thresholds in each ensemble system indicates that the WW200En better discerns between exceedance and non-exceedance of each IVT threshold (i.e., the WW200En ROCSS is largest; Fig. 11). However, dropsonde-based IVT ROC and ROCSS differences between each ensemble are marginal compared to 24-hour precipitation differences (Figs. 8a–c).

The spread-skill relationships for IVT (Fig. 12) yield broadly similar results for the three ensemble systems to the spread-skill relationship for precipitation (Fig. 7); however, the IVT results are notably noisier than those for precipitation, likely due to fewer forecast-observation pairs available using dropsonde data compared to gridded PRISM precipitation estimates. In general, forecast spread-skill curves for IVT from WW200En are to the right of the curves from the two global systems (i.e., more dispersive) for most lead times and spread thresholds (Fig. 12). At forecast day 1 (Fig. 12a), all three systems are underdispersive, with WW200En closest to the 1:1 line. Forecasts from all three ensembles are near the 1:1 line for most bins at forecast days 3 and 5 (Figs. 12b and c); however, WW200En becomes over-dispersive for IVT RMSEs greater than 100 kg m−1 s−1 compared to the two global systems at forecast day 3 (Fig. 12b). The 200-member ensemble is generally the most dispersive for most spread-error bins at forecast days 5 and 7 (Figs. 12c and d); however, all three ensembles are near the 1:1 line at day 5 and are generally left (along-to-right) of 1:1 for relatively small (large) errors at day 7. In summary, much like for precipitation forecasts over the western US, IVT forecast spread over the North Pacific appears to grow too slowly in all three ensemble systems, with spread growing most rapidly in WW200En.

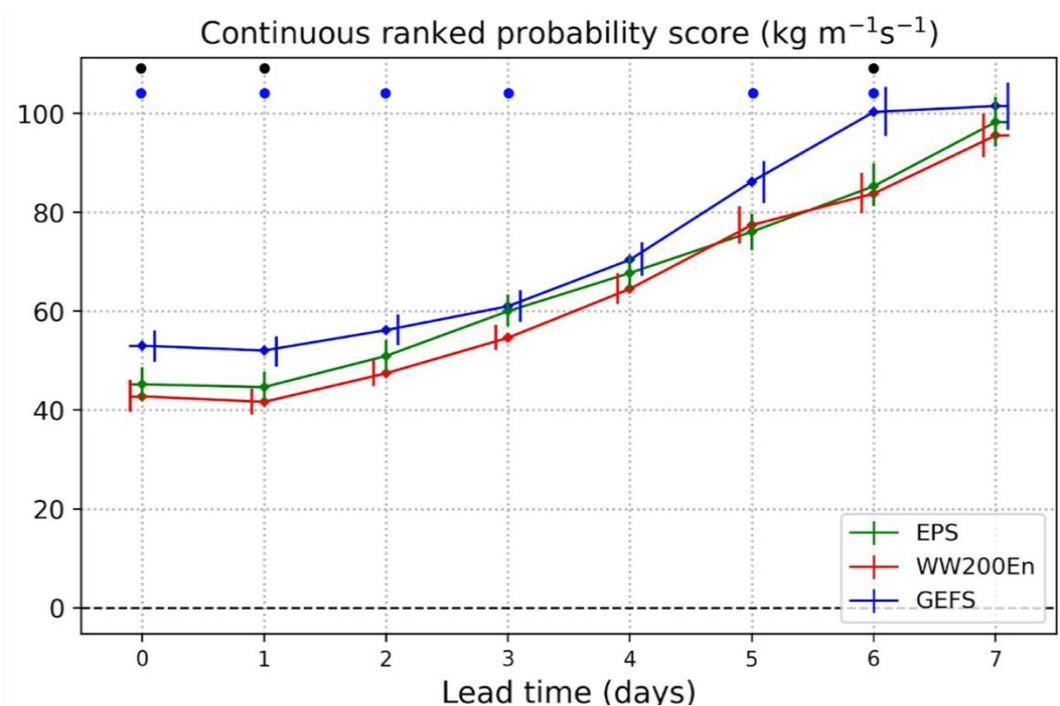

Fig. 9. Continuous ranked probability score (CRPS; smaller is better) for IVT with 90% bootstrap confidence intervals for all three ensembles. Results are aggregated over all dropsondes cases and plotted as a function of lead time. As a visual aid, a small offset was applied to the x-coordinates of the GEFS and WW200En bootstrap confidence intervals, and markers are placed along the upper x-axis denoting lead times where the differences between ensembles are statistically significant at the 90% confidence interval. Black markers refer to the difference between EPS and GEFS, blue markers refer to the difference between WW200En and GEFS, and green markers refer to the difference between WW200En and EPS.

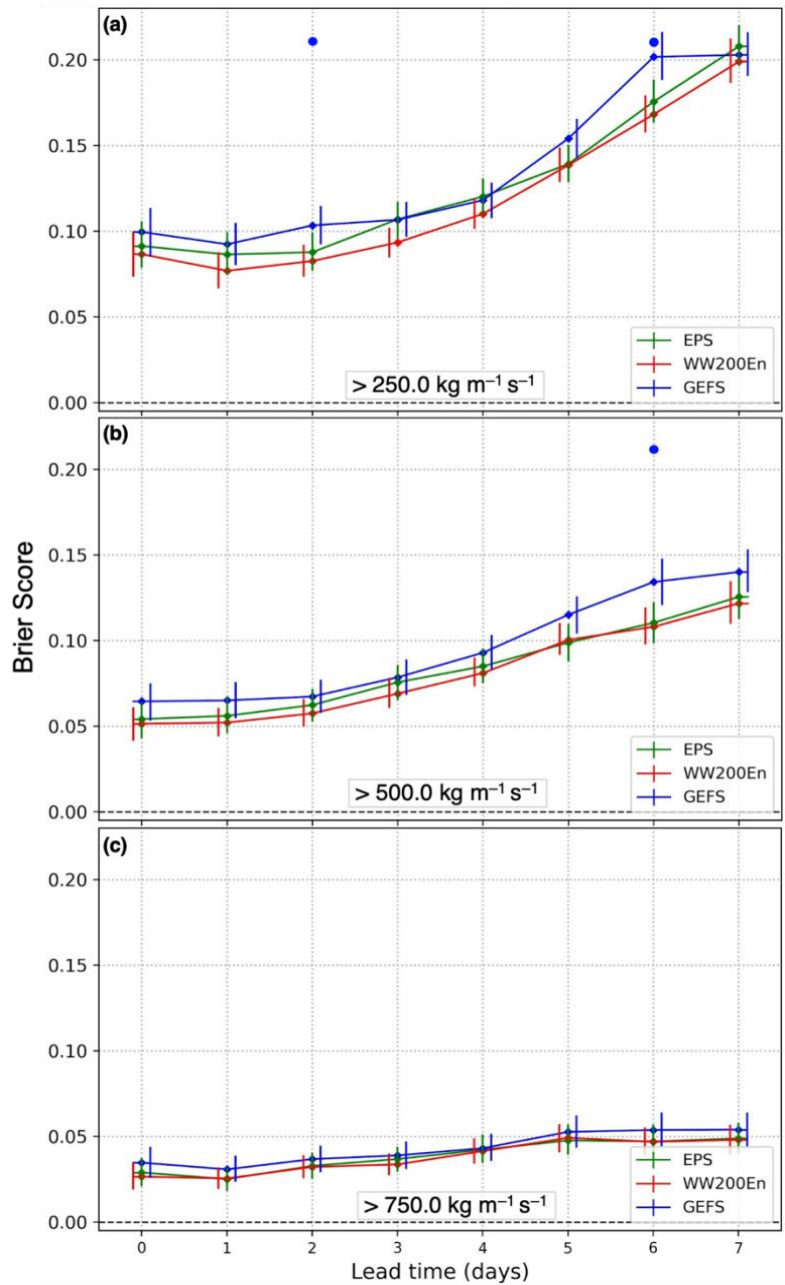

Fig. 10. Brier score (smaller is better) for each ensemble with 90% bootstrap confidence intervals for thresholds (a) > 250, (b) > 500, and (c) > 750 kg m-1 s-1. Results are aggregated over all dropsondes cases and plotted as a function of lead time. As a visual aid, a small offset was applied to the x-coordinates of the GEFS and WW200En bootstrap confidence intervals, and markers are placed along the upper x-axis denoting lead times where the differences between ensembles are statistically significant at the 90% confidence interval. Black markers refer to the difference between EPS and GEFS, blue markers refer to the difference between WW200En and GEFS, and green markers refer to the difference between WW200En and EPS.

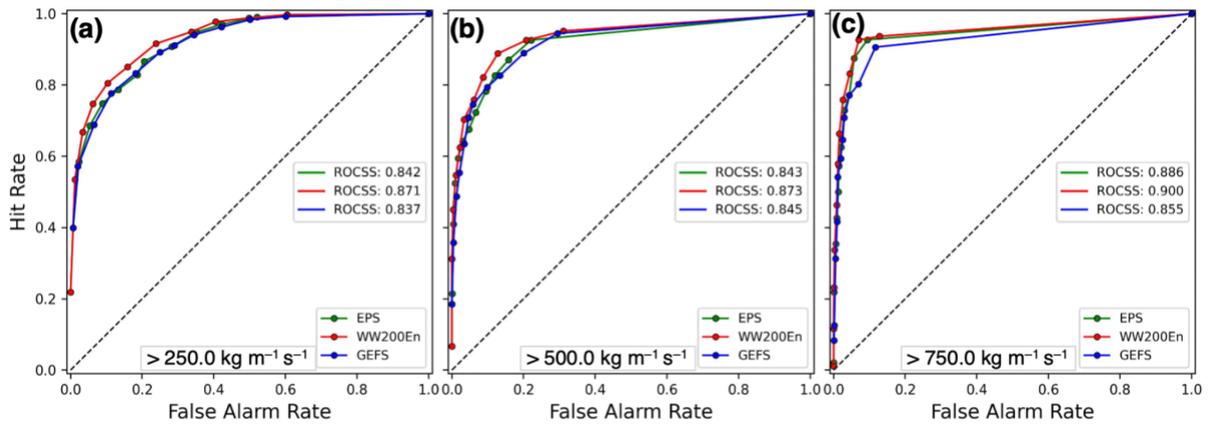

Fig. 11. As in Figure 8a–c, except for 72-hour IVT forecasts at thresholds (a) > 250, (b) > 500, (c) > 750 kg m-1 s-1.

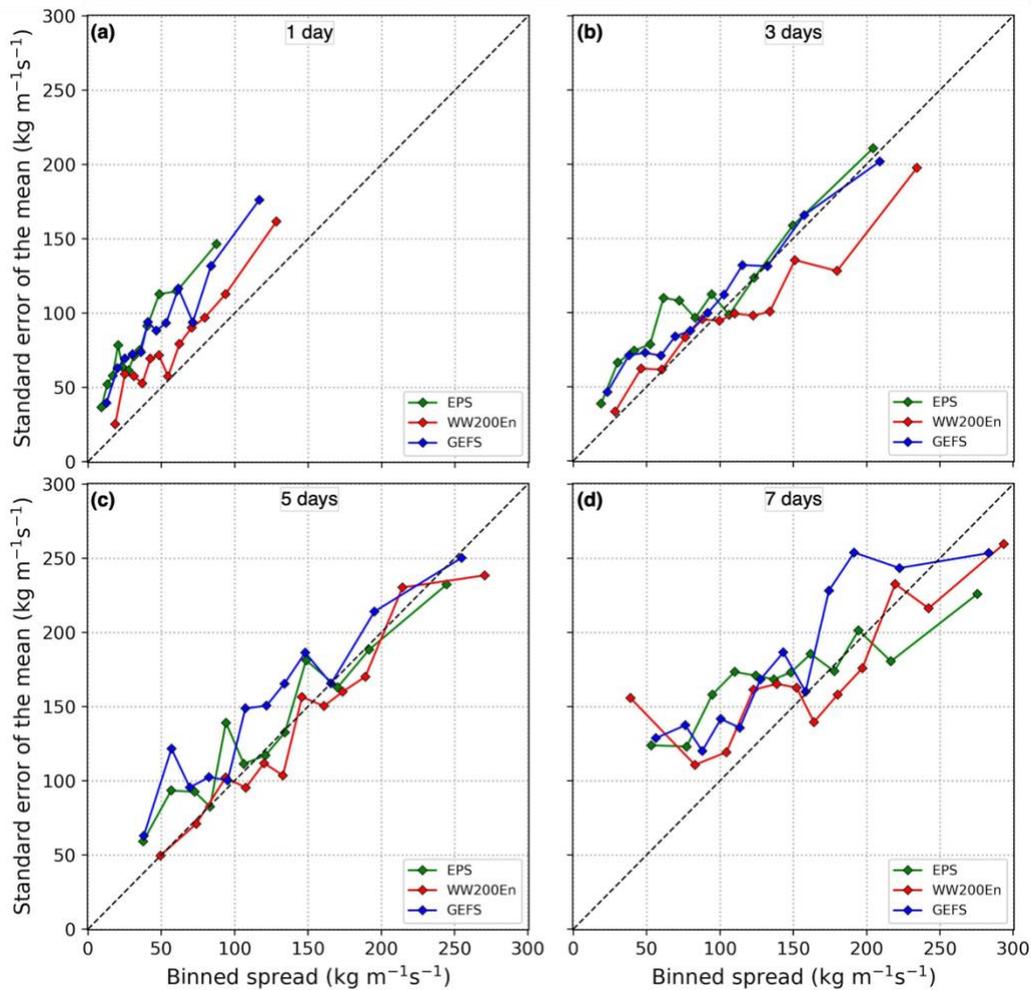

Fig. 12. Binned spread-skill diagrams for IVT for lead times +24-h (a), +72h (b), +120h (c), and +168h (d).

*c. Case studies*

During the 2022–2023 winter season, a series of nine ARs made landfall along the west coast of the US between 26 December and 17 January, resulting in record rain and snow accumulations throughout the state of California (DeFlorio et al. 2024). Two case studies during this exceptional AR period are examined here as a demonstration of the overall ensemble precipitation forecast differences between the WW200En, GEFS, and EPS described in section 5a. Three-day (i.e., 60–84 hour) forecasts of 24-h precipitation ending at 1200 UTC 31 December 2022 and ending at 1200 UTC 10 January are selected since these dates coincided with notable precipitation records, damaging floods, and forecast errors. For example, several precipitation records were set in the San Francisco Bay area on 31 December 2022 while landslides and damaging flooding occurred throughout the Coastal Mountains and Sierra foothills. The 1 through 5-day GFS deterministic forecasts (not shown) of the 31 December case consistently underestimated precipitation (errors as large as 100 mm within Cosumnes River watershed and surrounding Sierra foothills) as it missed the development of heavy precipitation in the Central Valley through the Sacramento/San Joaquin River delta (Fig. 13a). This was also the site of a levee break and resulted in devastating flooding. Around 10 January 2023, areas in the Transverse Ranges in Southern California received at least 100-200 mm of rain (Fig. 14a) and forecast models did not adequately capture these extremes. More than 70 rainfall records were set during this particular event (California Nevada River Forecast Center 2023).

During the December 2022 case (Fig. 13), the heaviest 24-h precipitation occurred along the Sierra Nevada mountains, exceeding 150–200 mm in the northern Sierras (Fig. 13a). Each of the three ensemble mean forecasts correctly identified the northern Sierras as the location of the heaviest 24-h precipitation; however, the total precipitation was underestimated by each ensemble mean forecast (Figs. 13b–d). The WW200En most closely resembles PRISM observations, producing a mean precipitation forecast of more than 100 mm over the central–northern Sierras (Fig. 13b), corresponding to an RMSE of 13.08 mm over the domain. This is consistent with differences between the ROC values for the prediction of 24-hour precipitation between each ensemble system over the 10-month period of study (Figs. 8a–c), where WW200En appears to better detect higher precipitation totals than the forecasts from the two global systems. By comparison, the GEFS (RMSE of 17.38 mm) and EPS (RMSE of 15.10) only produce maximum ensemble mean precipitation exceeding 75 mm and 100 mm, respectively, over very isolated locations in the northern Sierras (Figs. 13c and d). Additionally, EPS and WW200En mean forecasts both correctly identify a region of 24-h precipitation exceeding 50 mm along the northern California coast (Figs. 13a, b, and d).

Ensemble mean forecast skill is similar for the January 2023 case (Fig. 14), such that the locations of heaviest 24-h precipitation is generally well-represented by the ensemble mean (i.e., observations exceeding 150 mm in the southern and central Sierras and 200–250 mm in an elongated region encompassing much of Santa Barbara, Ventura, and Los Angeles Counties; Fig. 14a), but the mean forecast precipitation is too low. Moreover, WW200En generally over-predicts rainfall in the northern Sierras during this case, resulting in more similar RMSE scores between WW200En (20.58 mm; Fig. 14b) and EPS (20.75 mm; Fig. 14d) than during the December 2022 case.

Examining the probability of precipitation exceeding a specified threshold is another way to utilize an ensemble of forecasts beyond just the ensemble mean and spread and can provide insight on the likelihood of various scenarios at a given location (e.g., the probability of receiving any precipitation or the probability of receiving exceptionally heavy precipitation). Figure 15 depicts the probability of 24-h precipitation exceeding 90 mm for both the December 2022 and January 2023 cases from each of the three ensemble systems. Consistent with the relatively higher mean precipitation totals in WW200En, the latter is associated with the greatest exceedance probabilities for both cases of the three systems examined herein (Figs. 15a and d), particularly over the Sierras, as well as the highest single-case Brier Skill Scores. In the December 2022 case (Figs. 15a–c) each of the three systems correctly highlights a probability maximum over the central and northern Sierras, where observed precipitation exceeded 90 mm. Consistent with the behavior of the ensemble mean forecasts as well as the 10-month ROC differences (Figs. 8a–c), WW200En has the highest probability of exceedance over this region (70–80%), compared to the GEFS (30–40%) and EPS (50–60%) ensembles. Furthermore, both WW200En and EPS correctly indicate an increased likelihood of precipitation exceeding 90 mm over isolated areas of the north coast region, with WW200En better capturing terrain-scale variations in the distribution of exceedance probabilities (Figs. 15a and d).

Probability of exceedance results are similar for the January 2023 case, such that WW200En probabilities of in excess of 90 mm of precipitation exhibit terrain-scale variability over the central and southern Sierras and are between 90% and 100% over a portion of the region where at least 90 mm of precipitation was observed (Fig. 15d). While WW200En does have the highest single-case Brier Skill Score for this event (0.481), the relatively high probability of exceedance over the southern Sierras appears to be at the expense of erroneously high probabilities in the central and northern Sierras compared to the other systems, which also appear to extend the probability of 90 mm of precipitation too far north along the mountains (Figs. 15d–f). Finally, the 200-member ensemble appears to represent small-scale precipitation variability during this case better, such as along the north coast during the December 2022 event, as relatively high exceedance probabilities along the central coast are more isolated to the immediate coastal region than in the other two systems, more representative of 24-h rainfall observations (Figs. 15d–f).

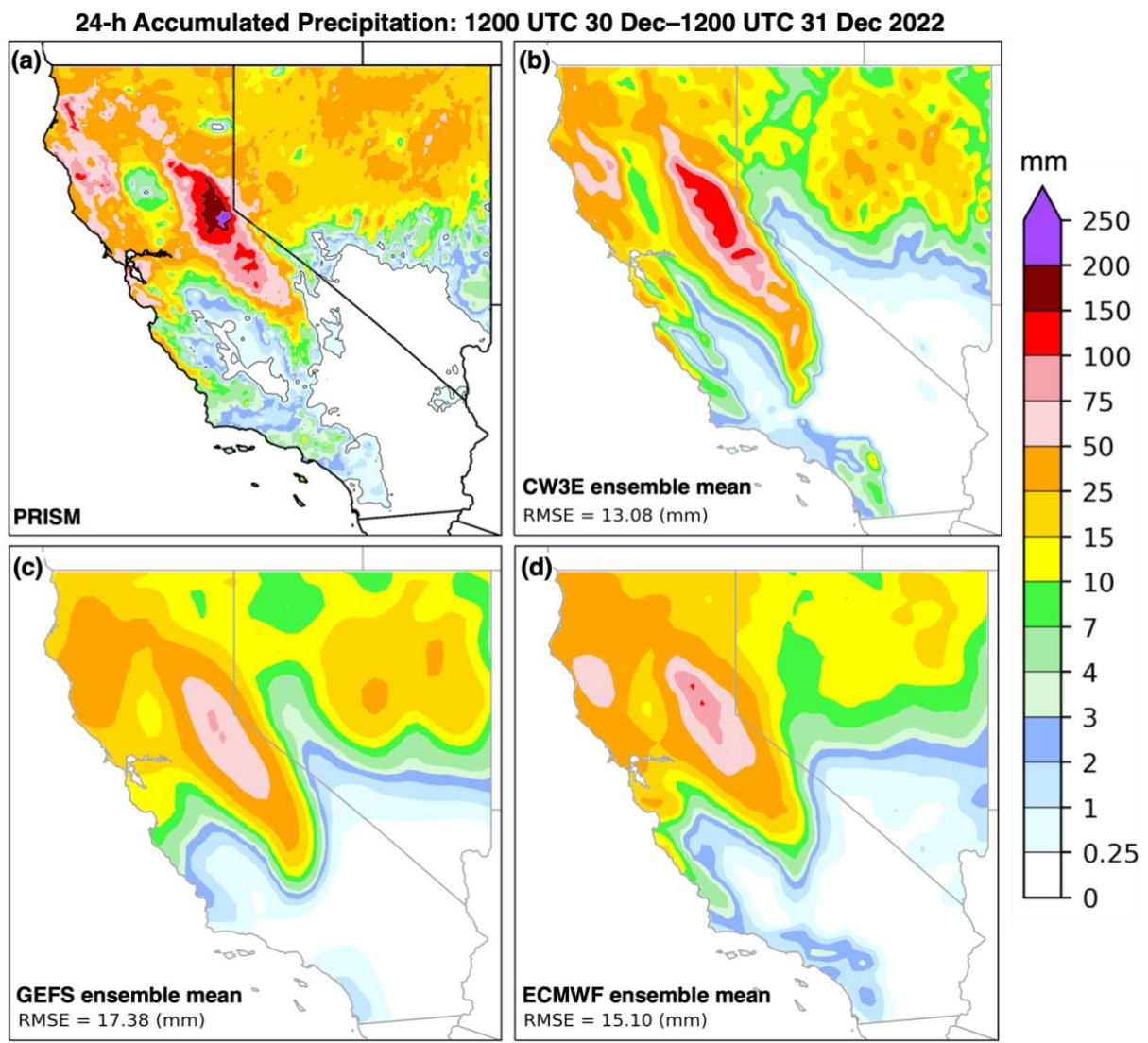

Fig. 13. The 24-h accumulated precipitation (mm) ending at 1200 UTC 31 December 2022 from (a) PRISM precipitation analysis and 3-day forecasts (lead time 60–84-hours) from (b)

the WW200En mean, (c) the GEFS ensemble mean, and (d) the EPS ensemble mean. Also shown is the RMSE for each ensemble calculated over the shown spatial domain and case.

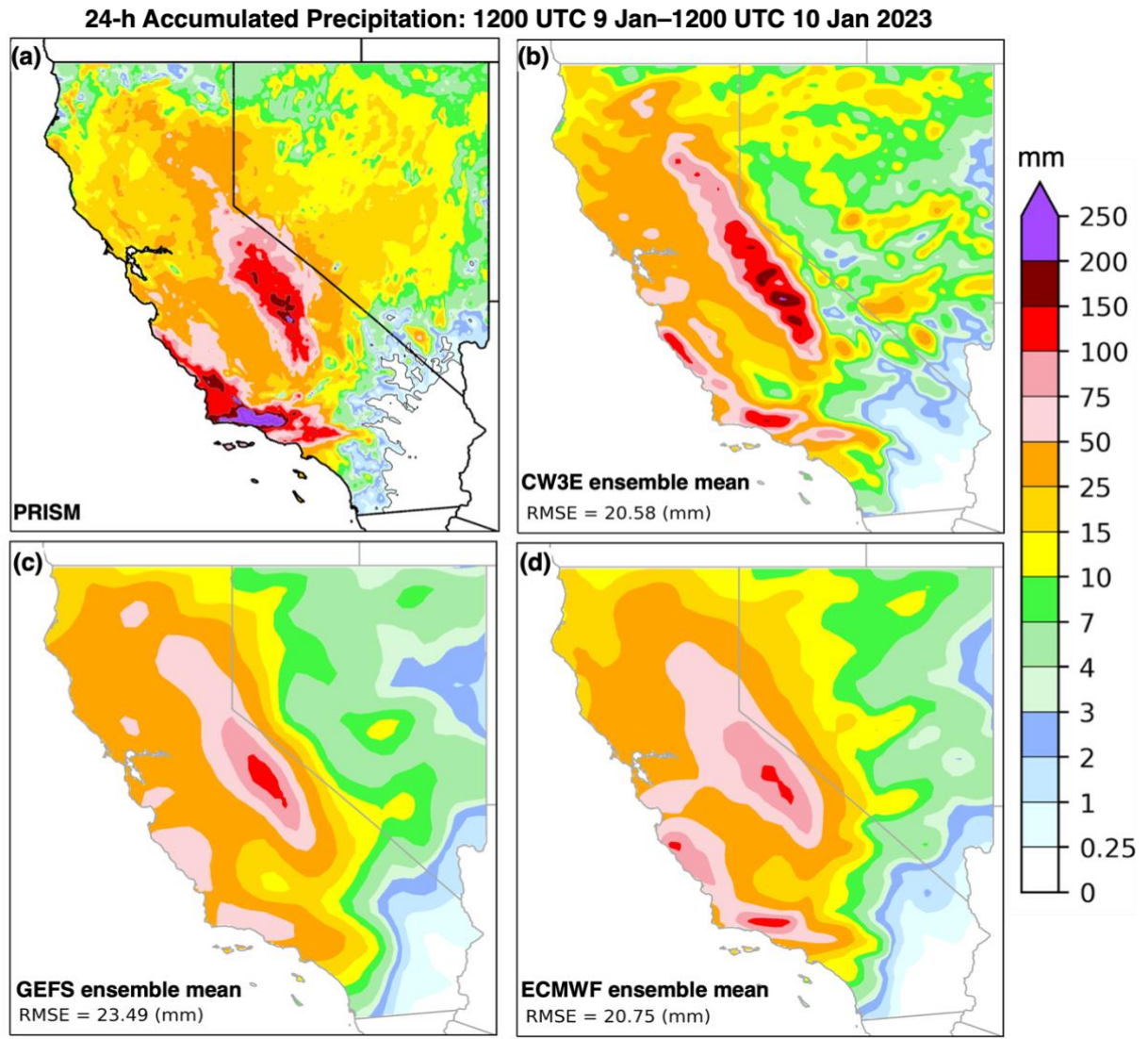

Fig. 14. As in Figure 13, except for 24-h precipitation ending at 1200 UTC 9 January 2023.

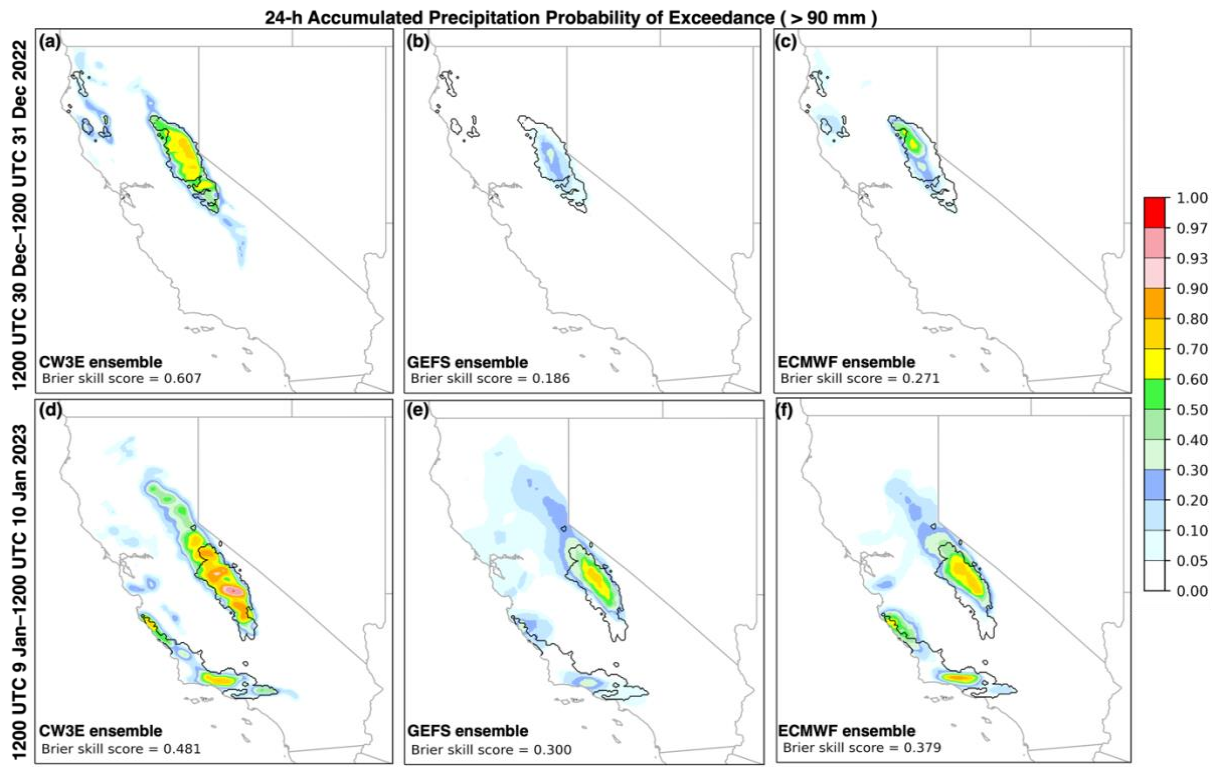

Fig. 15. 24-h 90 mm precipitation exceedance probability for 3-day forecast lead times (60–84 h) ending at (a–c) 1200 UTC 31 December 2022 and (d–f) 1200 UTC 10 January 2023 for WW200En (a and d), GEFS (b and e), and EPS (c and f). Black contours refer to the area over which the verifying PRISM 24-h precipitation analysis exceeds 90 mm. Also shown is the Brier skill score for each ensemble calculated over the shown spatial domain and case.

### d. *Sensitivity of precipitation forecast skill to the design of WW200En*

The WW200En consists of various simultaneous representations of errors in initial and boundary conditions and model formulation, including a multi-physics approach to represent uncertainty in physical parameterization schemes. While these approaches yield skillful probabilistic forecasts, as documented in the previous sections, it is possible that individual ensemble members may be prone to systematic errors related to their choice of physics parameterization schemes. This section evaluates probabilistic precipitation forecast skill in terms of the BSS and twCRPSS resulting from the exclusion of each microphysics, PBL, and cumulus parameterization scheme (one at a time) from the full 200-member ensemble, computed using the WW200En as the reference or "climatological" data. That is, the relative improvement or degradation of the Brier Score achieved by excluding members corresponding to specific parameterization schemes. Similar verification metrics were computed for IVT with respect to dropsonde observations; however, this analysis was generally inconclusive, likely due to the smaller sample size of discrete dropsonde observations compared to gridded PRISM precipitation data and is not discussed further. The resulting ensemble subsets contain 150 members when an individual PBL scheme is excluded, and 160 members when an individual microphysics or cumulus parameterization scheme is excluded.

One potential limitation of the physics-scheme exclusion approach used in the forthcoming section is the additional sources of uncertainty included in the 200-member ensemble, namely initial condition uncertainty and model uncertainty represented by SKEB. As a consequence, individual ensemble members will naturally differ in deterministic skill for reasons beyond their choice of physics parameterization. However, since this analysis includes daily forecasts over a 10-month period spanning two cool seasons (a total of 242 days), and the removal of individual parameterization schemes does not change the relative frequency of EPS and GEFS-based initial conditions (60% EPS, 40% GEFS), it is likely that any systematic impact of SKEB perturbations (which, for a single PBL, microphysics, or radiation scheme include 50, 40, or 40 unique time-evolving representations of the random noise field based on an initial, member-dependent random number seed) and the choice of initial and boundary conditions is minimal compared to that associated with physics parameterization.

1) Sensitivity to the selection of physics parameterizations

The largest change in precipitation forecast skill with respect to the full 200-member ensemble generally occurs at early forecast lead times (between forecast days 1–3) for experiments which exclude specific microphysics and PBL parameterization schemes. For the microphysics exclusion tests (Fig. 16a, d and Fig. 15a, d), the removal of members using the WSM6, or Purdue-Lin schemes yields the largest improvement in the ensemble forecasts, while excluding Thompson, WDM6, or Morrison members generally degrades the forecast skill. Changes in forecast skill measured by the BSS and twCRPSS, interpreted as the percent change in the BS or twCRPS with respect to the full 200-member ensemble, are larger for higher precipitation thresholds or percentiles (Fig. 16a, d and Fig. 15a, d). For example, exclusion of the 40 ensemble members which use the Thompson microphysics scheme worsens the > 90 mm BS by ~2–4% during forecast days 1–3, and similarly worsens the 99% twCRPS by ~2–3.5% during the same time (Fig. 16d and Fig. 17d). Conversely, exclusion of the 50 Purdue-Lin members improves the BS and twCRPS by ~1–3% and ~0.5–2.5%, respectively, at the same thresholds and during the same forecast lead times. At the higher precipitation thresholds, the change in precipitation skill due to removing specific microphysics members remains generally consistent for each subset during forecast days 3–6 (Fig. 16d and Fig. 17d).

Similar to the microphysics exclusion tests, the change in precipitation forecast skill by removing members associated with individual PBL parameterization schemes is largest during forecast days 1–3 and for higher precipitation thresholds (Fig. 16b, e and Fig. 17b, e). Exclusion of the 50 members using the MYJ PBL scheme consistently results in the largest degradation of precipitation forecast skill, measured by both BSS and twCRPSS, for each lead time and threshold included in Figure 16b, e and Figure 15b, e. Exclusion of ACM2 and YSU members generally yields smaller changes in forecast skill, particularly the twCRPSS (Figs. 17b and e), except for the > 25 mm BSS during forecast days 3–6 and day 6 90% twCRPSS, where excluding the YSU members yields the greatest BSS (~0.0025–0.005; Fig. 16b, e and Fig. 17b, e). Outside of these thresholds and times, the exclusion of the 50 MYNN PBL members results in the greatest improvement in precipitation forecast skill.

Unlike the microphysics and PBL exclusion tests, the change in precipitation forecast skill associated with removing individual cumulus parameterization schemes from the 200-member ensemble is relatively insensitive to forecast lead time and is also smaller than that observed in other exclusion tests. For example, the largest (removal of BMJ members) and smallest (removal of Kain-Fritsch members) > 90 mm BSS are only ~+0.011 and ~–0.016, respectively, which are 2–3 times smaller magnitude than the largest changes due to PBL or microphysics exclusion (Fig. 16). While the changes in precipitation skill scores are relatively small compared to other physics parameterizations, the exclusion of BMJ and Kain-Fritsch members is generally associated with the largest increase or decrease, respectively, in forecast skill for larger precipitation thresholds (Figs. 16 and 17).

Finally, Figure 18 depicts the precipitation BSS and twCRPSS for a 97-member ensemble constructed by excluding ensemble members which use any of the worst performing parameterization schemes associated with an individual parameterized process. Specifically, this subset consists of all members of the 200-member ensemble which do not use the WSM6 microphysics scheme, MYNN PBL scheme, or the BMJ cumulus scheme, which were excluded based on their overall performance as measured by BSS and twCRPSS at different thresholds. Except the > 25 mm BSS for forecast days 3–6, the skill score improvement for both metrics is larger than that for any single-scheme removal test for each lead time and threshold (Fig. 18). The low-threshold negative BSS for forecast days 4 and 5 is likely a consequence of the removal of members using the MYNN PBL and BMJ cumulus parameterizations since the exclusion of each of these schemes results in a negative > 25 mm BSS during later forecast lead times (Figs. 16b and c). Similar to the single-physics exclusion experiments, the largest improvement in precipitation forecast skill occurs during the first three forecast days, as skill scores decrease rapidly between days 1–3, and then stabilize, though still showing improvement compared to the full 200-member ensemble, from forecast days 3–6 (Fig. 18).

In summary, the precipitation forecast skill of WW200En ensemble is sensitive to the choice of physics parameterization schemes included in the multi-physics approach to represent model uncertainty, such that the inclusion or exclusion of members using specific physics configurations can degrade (or improve) probabilistic forecast skill, particularly during forecast days 1–3. Specifically, we find that the exclusion of members using Thompson microphysics results in the greatest loss of skill at all forecast lead times for particularly heavy precipitation, with smaller degradations associated with the removal of Morrison and WDM6 members (Figs. 16d–f). Notably, the Thompson, Morrison, and WDM6 microphysics schemes are each double-moment for at least two species, while those schemes whose exclusion improves precipitation forecast skill (Purdue-Lin and WSM6) are fully single moment schemes (Dudhia 2021). While a process-based evaluation is outside of the scope of the present study, these more computationally expensive double-moment schemes may more appropriately represent microphysical processes associated with orographic precipitation (e.g., l les et al. 2018) by predicting the number concentration of various microphysical species. The choice of PBL parameterization scheme appears to be of nearly comparable importance to microphysics parameterization for precipitation forecast skill over the western U.S., particularly at lead times of three days or less for relatively low precipitation thresholds (Figs. 16 and 17), while the choice of cumulus parameterization has a tertiary effect on precipitation skill. While both the microphysics and cumulus parameterization schemes directly modulate model precipitation, the overall larger influence of the choice of microphysics parameterization on forecast skill is likely due to greater resolved, grid-scale orographic precipitation over the coastal verification region (Fig. 1) compared to that associated with cumulus parameterization over the same region and the 10-month verification period. Moreover, the decreased, but non-zero, magnitude of the different contribution of the physical parameterizations beyond forecast day three may indicate that differences in the initial and boundary conditions between ensemble members may become a more significant source of ensemble diversity at longer forecast lead times compared to physics uncertainty.

2) SENSITIVITY TO ENSEMBLE SIZE EXCLUDING THE WORST-PERFORMING PHYSICS PARAMETERIZATIONS

The preceding section demonstrates that excluding members using specific parameterization schemes from the WW200En generally results in more skillful precipitation forecasts over a 10-month study period. Before considering the addition of 103 members to the 97-member subset, the response of precipitation forecast skill to the size of the ensemble subset should be examined. For example, Mullen and Buizza (2002) concluded that, for the 51-member ECMWF ensemble at their time of writing, increasing the number of ensemble members would yield a greater improvement in probabilistic precipitation forecast skill compared to computationally similar higher-resolution, fewer-member ensembles. In more recent research using the convection-permitting AROME ensemble at Météo France, Raynaud and Bouttier (2017) similarly determined that a 34-member, 2.5 km grid spacing ensemble provided greater precipitation forecast skill compared to 12-member ensembles with 1.3 km and 2.5 km grid spacing, particularly noting that improvements due to increased ensemble size become more prominent with increasing forecast lead time and increasing rainfall rate.

Consistent with previous research, precipitation forecast skill quantified using the BSS and twCRPSS with respect to the full WW200En tends to increase with increasing ensemble size, particularly for larger rainfall rates and forecast lead times. Randomly subsetting the 97-member ensemble in intervals of 10-members from 10 to 90, and finally the full 97 members, and comparing precipitation forecast skill to the full 200-member ensemble, it is evident that ensembles of 20 members or fewer results in largely degraded forecast skill, regardless of forecast lead time (Fig. 19). For shorter lead times (e.g., forecast days 1 and 3; Figs. 19a–d), subsets of 30 members or more generally result in comparable or improved performance relative to the full 200-member ensemble at thresholds greater than 0.254 mm (i.e., probability of any precipitation). Interestingly, the skill of predicting the probability of any precipitation is degraded with respect to the 200-member ensemble for all ensemble sizes and lead times in Figure 19. For longer forecast lead times (e.g., forecast day 3; Figs. 17c-d), ensemble subset skill generally increases from 50–80 members, most notably for the >90 mm BSS and 99th percentile twCRPSS (yellow and blue lines in Fig. 19e and 19f, respectively). Moreover, both the high-threshold BSS and twCRPSS continue to marginally increase between 80 and 97 members, indicating that using the entire 97-member ensemble is beneficial compared to further subsetting the best-performing members, and perhaps further suggesting that increasing the size of the best-performing subset beyond 97-members may yield further improvements in precipitation forecast skill at lead times of three days or more. A similar analysis of forecast skill as a function of ensemble subset size was performed for the whole 200-member ensemble, with little improvement beyond ~80 members (not shown), suggesting that the best-performing 97-member ensemble may be relatively more sensitive to ensemble size.

Finally, precipitation forecasts from the 97-member ensemble and its subsets are compared to those from the GEFS and EPS (i.e., rightmost two abscissa points in Fig. 19). Similar to the evaluation of the BSS with respect to climatology for the three full ensemble systems (Fig. 4), the relative performance of subsets of the 97 best-performing members with respect to the two global ensembles varies depending on the chosen precipitation threshold and forecast lead time. For the prediction of the probability of any precipitation, the BSS of EPS and GEFS forecasts (computed with respect to the 200-member ensemble) are smaller than for all subset sizes and lead times, except for the 10-member subset at forecast day 6 (Fig. 19e). Moreover, only EPS 50 mm BSS is larger than the corresponding BSS for a subset of comparable size (~50 members) at forecast day 3, and both the GEFS and EPS BSS for 75 and 90 mm are larger than the corresponding BSS for comparably sized subsets (~30 members and ~50 members, respectively) at forecast day 6 (Figs. 19c and e). The only threshold and lead time at which the two global systems have a larger BSS than any subset of the 97-member ensemble is for precipitation exceeding 90 mm at forecast day 6 (Fig. 19e), while for the percentile thresholds, the EPS twCRPSS is largest for the 95th (99th) percentile at forecast days 3 and 6 (day 6), and the GEFS twCRPSS is only larger than all subsets for the 99th percentile at forecast day 6 (Fig. 19f). Overall, comparing precipitation forecast skill metrics from subsets of the 97-member best-performing ensemble with those from the GEFS and EPS yields qualitatively similar results to the those derived from comparisons of the two global ensembles with the WW200En (e.g., Fig. 4 and section 5a). Specifically, instances where the subset BSS is less than that associated with EPS or GEFS for comparable ensemble sizes (i.e., day 3 for EPS at 50 mm, day 6 for EPS and GEFS at 75 and 90 mm; Figs. 4d, g, i, and Figs. 19c, f) are consistent, suggesting that conclusions drawn by comparing the 200-member WW200En to the 51-member EPS and 31-member GEFS are not entirely dependent on ensemble size.

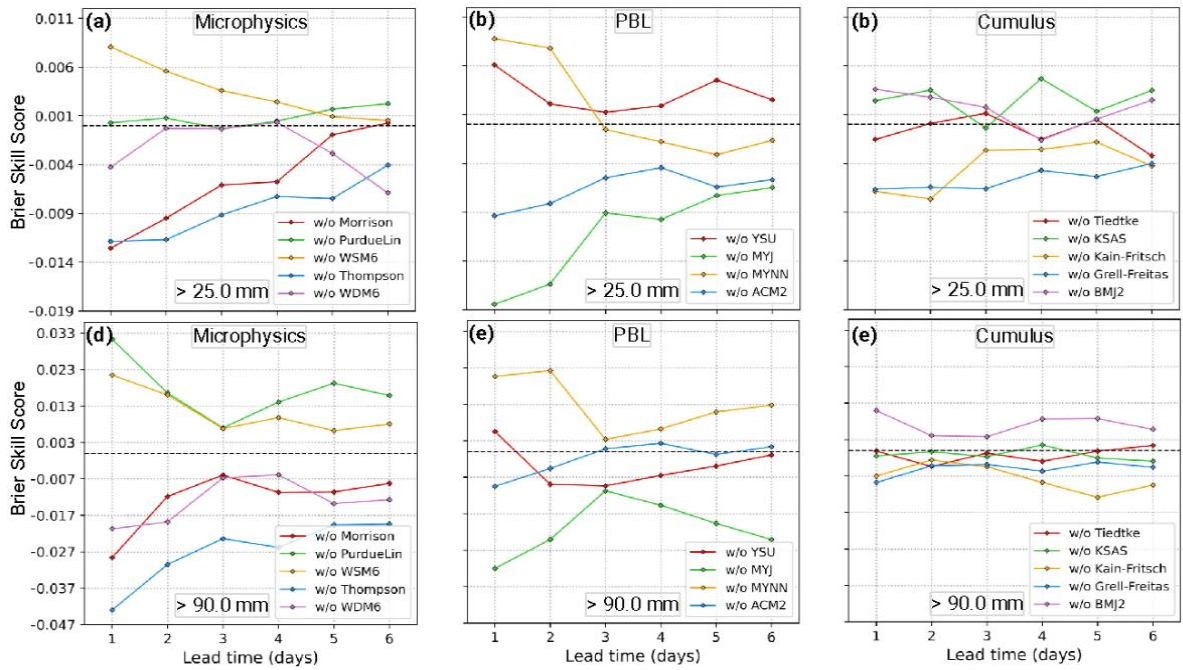

Fig. 16. Brier Skill Scores (BSS) for 24-h precipitation computed from ensemble members excluding individual parameterization schemes as indicated by legend labels for microphysics schemes (a and d), PBL schemes (b and e), and cumulus schemes (d and f) computed using the full 200-member ensemble as the reference forecast. Panels (a–c) and (d–f) refer to 25 mm and 90 mm 24-h precipitation thresholds, respectively.

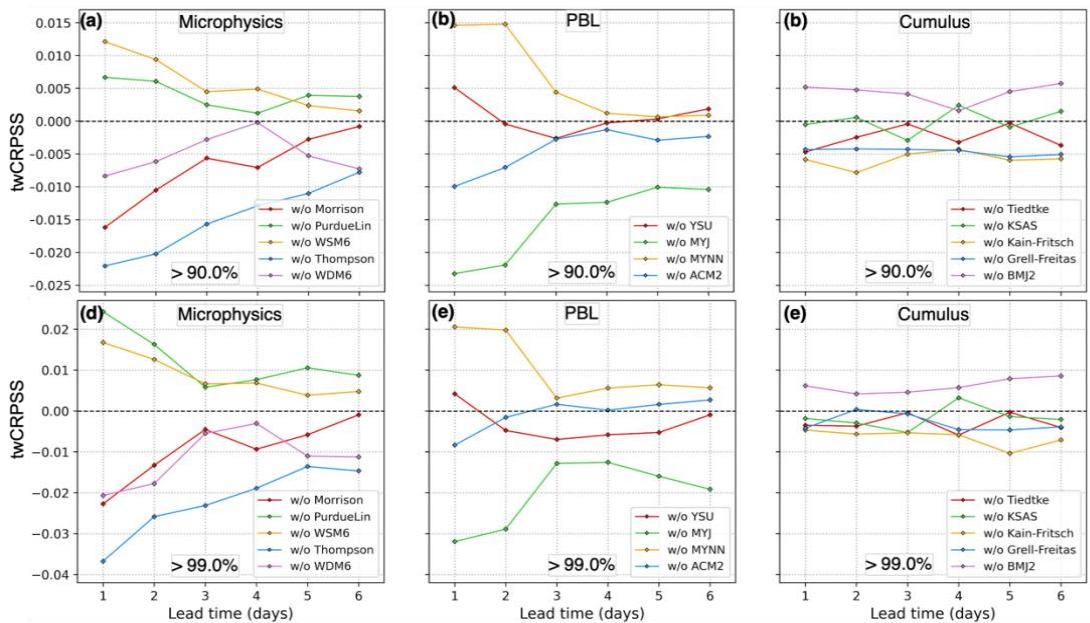

Fig. 17. As in Figure 16, except for (a–c) 90th percentile and (d–f) 99th percentile threshold-weighted continuous-ranked probability skill score.

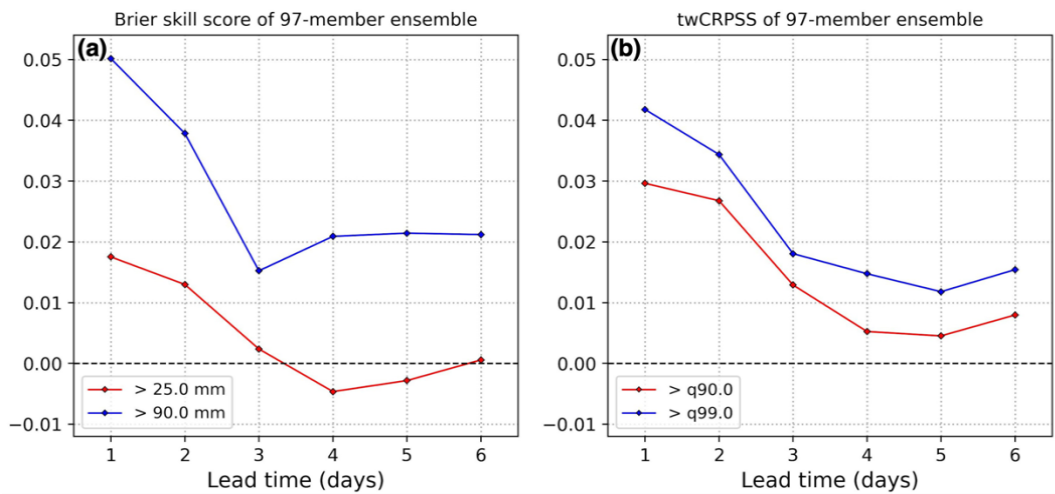

Fig. 18. Brier Skill Score (BSS; a) and threshold-weighted continuous-ranked probability skill score (twCRPSS; b) for 24-h precipitation computed from ensemble members excluding those using the WSM6 microphysics scheme, MYNN PBL scheme, and BMJ cumulus scheme, for a total of 97 remaining members. Skill scores are computed using the full 200-member ensemble as the reference forecast. Lines refer to four 24-h precipitation thresholds: (a; red) > 25 mm, (a; blue) > 90 mm, (b; red) > 90th percentile, and (b; blue) > 99th percentile.

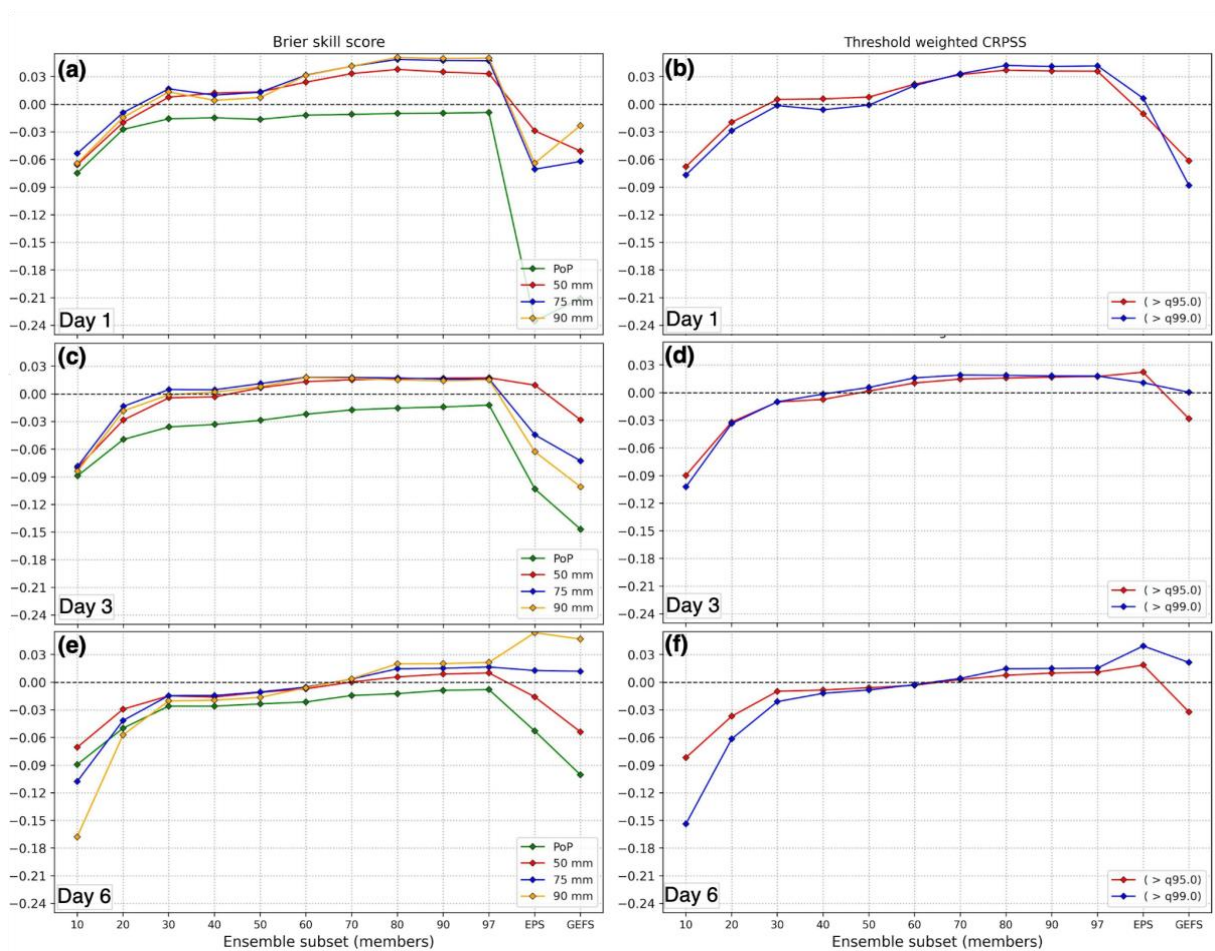

Fig. 19. Brier Skill Score (BSS; a, c, and e) and threshold-weighted continuous-ranked probability skill score (twCRPSS; b, d, and f) for 24-h precipitation computed from EPS, GEFS, and N = 10, 20, …, 90, 97 randomly-selected (x-axis) ensemble members from the 97-member subset that excludes those using the WSM6 microphysics scheme, MYNN PBL scheme, and BMJ cumulus scheme computed using the full 200-member ensemble as the reference forecast; (a and b) refer to a 1-day forecast lead time, (c and d) to 3-days, and (e and f) to 6-days. BSS from the full EPS and GEFS ensembles are included at the rightmost two x-axis points for reference.

## 6. Discussion and Concluding Remarks

Since 2015, the Center for Western Weather and Water Extremes (CW3E) has generated high-resolution, near real-time deterministic forecasts of atmospheric rivers (ARs) and their associated precipitation over the western U.S. with a tailored version of the Weather Research and Forecast (WRF) model, called West-WRF. Beginning in 2021, CW3E has also generated probabilistic forecast products based on a 9-km resolution 200-member ensemble based on West-WRF (WW200En), which accounts for uncertainty in the initial and boundary conditions, parameterized physics, and subgrid-scale numerics. The present study describes the configuration of the WW200En and examines its predictive skill over a 10-month period of study in comparison to probabilistic forecasts from global forecast systems, namely the US National Center for Environmental Prediction (NCEP)'s Global Ensemble Forecast System (GEFS) and the European Centre for Medium-Range Weather Forecasts (ECMWF)'s Ensemble Prediction System (EPS). Importantly, both the deterministic West-WRF and WW200En are specifically designed to represent AR-based extreme precipitation events, using analyses and forecasts from the global systems (GEFS, EPS) as initial and boundary conditions and as a baseline from which to provide additional skill. These modeling systems are utilized by a variety of stakeholders, including those from local agencies, the California Department of Water Resources, and the U.S. Army Corps of Engineers involved in the Forecast Informed Reservoir Operations (Talbot et al. 2019) and the AR Reconnaissance campaign (Ralph et al. 2020).

Indeed, WW200En precipitation forecast skill and spread compares favorably to these leading global systems, demonstrating improved probabilistic skill with respect to the GEFS at nearly all thresholds and lead times, and at least comparable skill with respect to the EPS. At larger 24-h accumulated rainfall thresholds (e.g., greater than 75, 80, or 90 mm), WW200En typically exhibits the best forecast skill metrics (e.g., BSS and its components, ROC, and Reliability diagrams; Figs. 4–6, and 8), consistent with previous research indicating that probabilistic precipitation forecasts improve with increasing ensemble size for particularly large precipitation rates (Mullen and Buizza 2002; Raynaud and Bouttier 2017). Additionally, WW200En exhibits a better ability to quantify the prediction uncertainty compared to GEFS and EPS as shown by binned spread-skill diagrams compiled for different lead times (Fig. 7), as well as an improved overall reliability and resolution of the probabilistic prediction. The results for integrated vapor transport (IVT) forecast verification across the three ensemble systems are qualitatively similar to those for precipitation forecasts (i.e., WW200En and EPS are comparably skillful, and generally perform better than the GEFS). In summary, we find that the 9-km 200-member regional ensemble provides ~0.5–2 days of additional predictability compared to the two global systems described herein, depending on forecast metric or threshold analyzed. While outside of the scope of the present study, which provides an introduction of CW3E's WW200En and an overall evaluation of its forecast skill, it would be of future interest to examine the performance of the 200-member ensemble under various, potentially error-prone meso–synoptic-scale flow configurations, e.g., as the North Pacific blocking (Moore et al. 2021), or the development of mesoscale frontal waves (Martin et al. 2019; Hecht et al. 2022). Further evaluation of the present composition of WW200En ensemble may yield insight into potential avenues for its future development.

By comparing precipitation forecast skill of 150–160-member ensemble subsets, each subset constructed by excluding members containing a specified parameterization scheme, to the full 200-member ensemble, this study finds that the WW200En is particularly sensitive to the choice of microphysics and PBL parameterization schemes, and relatively less sensitive to the choice of cumulus scheme, particularly during forecast days 0–3. Specifically, we find that a 97-member ensemble composed of members that do not use any of the WRF Single-Moment 6-Class (WSM6) microphysics, Mellor-Yamada-Nakanishi-Niino (MYNN) PBL, or Betts-Miller-Janjic (BMJ) cumulus schemes generally outperforms the 200-member ensemble at all lead times for high precipitation thresholds. Additional research, such as a further process-oriented examination of the case studies described in section 5c, is warranted to better understand why certain physics parameterization schemes perform better or worse than others in the WW200En.

Precipitation forecast skill generally increases with increasingly large subsets of this "best-performing" 97-member ensemble for high precipitation thresholds, suggesting that constructing additional members based on the 97-member configuration may yield further improvements to forecast skill at high precipitation rates and long forecast lead times. The redistribution of 103 members worth of computational resources could consist in additional or alternative stochastic schemes to further explore the uncertainty stemming from the physics packages adopted in the best-performing 97 members. These could include stochastically perturbed parameterization tendencies, stochastically perturbed parameterizations, or physics-based stochastic perturbations schemes (Bouttier et al. 2012; Kober and Craig 2016; Zhou et al. 2022). Additionally, results from ongoing research comparing West-WRF ensemble output at 9-km vs 3-km horizontal grid spacing indicates that there are clear benefits with finer grid spacing, particularly up to 2 or 3 days, although an increased resolution (horizontal, vertical, or both) will reduce the size of the ensemble given the same computational resources, which could degrade the performance for the prediction of extreme events. Given the importance of orographic precipitation process over the region of focus for this study, ongoing research at CW3E is also evaluating strategies for the perturbations of model terrain elevation, similarly to Li et al. (2021).

One of the challenges associated with multiphysics ensemble systems is that ensemble members cannot be considered equally-likely outcomes due to varying biases and uncertainties between physical parameterization schemes. The sensitivity experiments described in section 5d underscore this point, such that the exclusion of various physical parameterization schemes notably degrades (or improves) the ensemble performance. These experiments can be further interpreted as a kind of postprocessing, or calibration, of the raw WW200En output, where the excluded members are calibrated to have zero-weight. Further research evaluating the performance of postprocessed WW200En forecasts is ongoing. For example, Ghazvinian et al. (2024) used a novel deep learning approach to postprocess 24-hour accumulated precipitation forecasts derived from WW200En, demonstrating that calibrating raw ensemble forecasts led to substantial forecast improvements, especially for heavy to extreme precipitation events. While the 2021–2022 and 2022–2023 periods of study encompass frozen versions of the global operational systems analyzed in this study, we note that, at the time of writing, the horizontal resolution of the ECMWF EPS has increased to ~9 km following the upgrade to CY48R1 (ECMWF 2023) in June 2023. Continued research is planned to evaluate the performance of the WW200En relative to state-of-the-art global operational systems.

*Acknowledgments*

This study and the Comet supercomputer, which is managed by the San Diego Supercomputing Center, were supported and made available by the Atmospheric River Program Phase III and IV awarded by California Department of Water Resources (contract numbers 4600014294 and 4600014942) and the U.S. Army Corps of Engineers (USACE) Forecast Informed Reservoir Operations Awards (USACE W912HZ1920023 and USACE W912HZ-15-2-0019). We are thankful to Michael Brunke, Brian Kawzenuk, Aneesh Subramanian, and Ryan Torn for useful discussions on the design of the Center for Western Weather and Water Extremes' West-WRF 200-member ensemble. Aneesh Subramanian and Ryan Torn also provided valuable comments to the manuscript. We would like to extend our sincere gratitude to the anonymous reviewers for their insightful comments and constructive feedback, which have greatly improved the quality of this manuscript.

*Data Availability Statement*

The PRISM analysis data are publicly available from the Oregon State University at https://www.prism.oregonstate.edu/. The West-WRF output is too large to be publicly archived with available resources. Access to model output, along with documentation and methods used to support this study, are available from author Daniel Steinhoff at CW3E/UCSD (dsteinhoff@ucsd.edu). NCEP GEFS operational dataset can be downloaded from https://registry.opendata.aws/noaa-gefs/. Archived ECMWF ensemble forecast data are available via https://www.ecmwf.int/en/computing/software/ecmwf-web-api. Near real-time ECMWF operational ensemble forecasts were obtained based on an agreement between CW3E and the ECMWF.

File generated with AMS Word template 2.0

aircraft and regional mesonet observations of orographic precipitation and its forcing. *J Hydrometeor.*, **19,** 1097–1113, https://doi.org/10.1175/JHM-D-17-0098.1.

Martin, A. C., F. M. Ralph, A. Wilson, L. DeHaan, and B. Kawzenuk, 2019: Rapid cyclogenesis from a mesoscale frontal wave on an atmospheric river: impacts on forecast skill and predictability during atmospheric river landfall. J. Hydrometeor., 20, 1779–1794, https://doi.org/10.1175/JHM-D-18-0239.1.

Marzban, C., R. Tardif, S. Sandgathe, and N. Hryniw, 2019: A methodology for sensitivity analysis of spatial features in forecasts: the stochastic kinetic energy backscatter scheme. *Meteor. Appl.*, **26,** 454–467, https://doi.org/10.1002/MET.1775.

Mcguire, L. A., F. K. Rengers, N. Oakley, J. W. Kean, D. M. Staley, H. Tang, M. De Orla-Barile, and A. M. Youberg, 2021: Time since burning and rainfall characteristics impact post-fire debris-flow initiation and magnitude. *Environ. Eng. Geosci.*, **27**, 43–56, https://doi.org/10.2113/EEG-D-20-00029.

McTaggart-Cowan, R., and Coauthors, 2022a: Using stochastically perturbed parameterizations to represent model uncertainty. Part I: implementation and parameter sensitivity. *Mon. Wea. Rev.*, **150,** 2829–2858, https://doi.org/10.1175/MWR-D-21-0315.1.

——, L. Separovic, M. Charron, X. Deng, N. Gagnon, P. L. Houtekamer, and A. Patoine, 2022b: using stochastically perturbed parameterizations to represent model uncertainty. Part II: comparison with existing techniques in an operational ensemble. *Mon. Wea. Rev.*, **150,** 2859–2882, https://doi.org/10.1175/MWR-D-21-0316.1.

Mesinger, F., 1993: Forecasting upper tropospheric turbulence within the framework of the Mellor-Yamada 2.5 closure. Research Activities in Atmospheric and Oceanic Modeling Rep. **18**, WMO, 4.28–4.29.

Molteni, F., R. Buizza, T. N. Palmer, and T. Petroliagis, 1996: The ECMWF Ensemble Prediction System: methodology and validation. *Quart. J. Roy. Meteor. Soc.*, **122,** 73–119, https://doi.org/10.1002/QJ.49712252905.

Morales, A., H. Morrison, and D. J. Posselt, 2018: Orographic precipitation response to microphysical parameter perturbations for idealized moist nearly neutral flow. J. Atmos. Sci., **75**, 1933–1953, https://doi.org/10.1175/JAS-D-17-0389.1.

File generated with AMS Word template 2.0

File generated with AMS Word template 2.0

——, and F. Bouttier, 2017: The impact of horizontal resolution and ensemble size for convective-scale probabilistic forecasts. *Quart. J. Roy. Meteor. Soc.*, **143**, 3037–3047, https://doi.org/10.1002/QJ.3159.

San Diego Super Computer Center, 2021a: Comet user guide. UC San Diego, Accessed 29 October 2024, https://www.sdsc.edu/support/user_guides/comet.html.

——, 2021b: Retired resources. UC San Diego, Accessed 29 October 2024, https://www.sdsc.edu/support/user_guides/retired.html.

Schwartz, C. S., G. S. Romine, and D. C. Dowell, 2021: Toward unifying short-term and next-day convection-allowing ensemble forecast systems with a continuously cycling 3-km ensemble Kalman filter over the entire conterminous United States. *Wea. Forecasting*, **36,** 379–405, https://doi.org/10.1175/WAF-D-20-0110.1.

Skamarock, C., and Coauthors, 2019: A description of the Advanced Research WRF model version 4. https://doi.org/10.5065/1DFH-6P97.

Song, M., and Coauthors, 2024: Non-crossing quantile regression neural network as a calibration tool for ensemble weather forecasts. *Adv. Atmos. Sci.*, **41,** 1417–1437, https://doi.org/10.1007/S00376-023-3184-5/METRICS.

Stensrud, D. J., J.-W. Bao, and T. T. Warner, 2000: Using initial condition and model physics perturbations in short-range ensemble simulations of mesoscale convective systems. *Mon. Wea. Rev.*, **128,** 2077–2107, https://doi.org/10.1175/1520-0493(2000)128.

Stone, R. E., C. A. Reynolds, J. D. Doyle, R. H. Langland, N. L. Baker, D. A. Lavers, and F. M. Ralph, 2020: Atmospheric River Reconnaissance observation impact in the Navy Global Forecast System. *Mon. Wea. Rev.*, **148,** 763–782, https://doi.org/10.1175/MWR-D-19-0101.1.

Sun, W., Z. Liu, C. A. Davis, F. M. Ralph, L. D. Monache, and M. Zheng, 2022: Impacts of dropsonde and satellite observations on the forecasts of two atmospheric-river-related heavy rainfall events. *Atmos. Res.*, **278,** 106327, https://doi.org/10.1016/J.ATMOSRES.2022.106327.

Talbot, C. A., M. Ralph, J. Jasperse, and J. Forbis, 2017: Forecast-Informed Reservoir Operations: lessons learned from a multi-agency collaborative research and operations effort to improve flood risk management, water supply and environmental benefits. *AGU Fall Meeting*, New Orleans, LA, H11J-1338.

——, Ralph F. M., and Jasperse J., 2019: Forecast-informed reservoir operations: lessons learned from a multi-agency joint research and operations effort. *Federal Interagency Sedimentation and Hydrologic Modeling Conf.,* Reno, NV.

Teixiera, J., and C. A. Reynolds, 2008: Stochastic nature of physical parameterizations in ensemble prediction: a stochastic convection approach. *Mon. Wea. Rev.*, **136,** 483–496, https://doi.org/10.1175/2007MWR1870.1.

Thomas, S. J., J. P. Hacker, and R. B. Stull, 2002: An ensemble analysis of forecast errors related to floating point performance. *Wea.Forecasting*, **17,** 898–906.

Thompson, G., P. R. Field, R. M. Rasmussen, and W. D. Hall, 2008: Explicit forecasts of winter precipitation using an improved bulk microphysics scheme. Part II: implementation of a new snow parameterization. *Mon. Wea. Rev*. **136,** 5095–5115, https://doi.org/10.1175/2008MWR2387.1.

Toth, Z., and E. Kalnay, 1997: Ensemble forecasting at NCEP and the breeding method. *Mon. Wea. Rev.*, **125,** 3297–3319, https://doi.org/10.1175/1520-0493(1997)125.

Waliser, D., and B. Guan, 2017: Extreme winds and precipitation during landfall of atmospheric rivers. *Nat. Geosci.,* **10,** 179–183, https://doi.org/10.1038/NGEO2894.

Wang, X., and C. H. Bishop, 2003: A comparison of breeding and ensemble transform Kalman filter ensemble forecast schemes. *J. Atmos. Sci.,* **60,** 1140–1158, https://doi.org/10.1175/1520-0469(2003)060<1140:ACOBAE>2.0.CO;2.

Wilks, D., 2011: *Statistical methods in the atmospheric sciences*.

Yang, S. C., T. H. Yang, Y. C. Chang, C. H. Chen, M. Y. Lin, J. Y. Ho, and K. T. Lee, 2020: Development of a hydrological ensemble prediction system to assist with decision-making for floods during typhoons. *Sustainability,* **12,** 4258, https://doi.org/10.3390/SU12104258.

Zhang, C., and Y. Wang, 2017: Projected future changes of tropical cyclone activity over the western north and south Pacific in a 20-km-mesh regional climate model. *J. Clim.*, **30,** 5923–5941, https://doi.org/10.1175/JCLI-D-16-0597.1.
52File generated with AMS Word template 2.0